\documentclass[12pt]{article}
\usepackage{amssymb,amsmath,latexsym,verbatim,graphicx,young}

\usepackage[usenames]{color}

\usepackage{cite}

\def\6#1{{\underline{#1}}}
\def\m6#1{{\underline{#1}\,}}

\newdimen\Tdim
\def\ispan{{\setbox0=\hbox{i}%
\Tdim\ht0\advance\Tdim\dp0\rule[-\dp0]{0pt}{\Tdim}}}
\def\jspan{{\setbox0=\hbox{j}%
\Tdim\ht0\advance\Tdim\dp0\rule[-\dp0]{0pt}{\Tdim}}}
\def\Tspan#1{{\setbox0=\hbox{#1}%
\Tdim\ht0\advance\Tdim\dp0\advance\Tdim.55ex\rule[-\dp0]{0pt}{\Tdim}\box0}}
\def\be{\begin{eqnarray}}
\def\ben{\begin{eqnarray*}}
\def\ee{\end{eqnarray}}
\def\een{\end{eqnarray*}}

\def\D{\mathcal{D}}

\def\=:{=\hspace{-.7em}\raisebox{1.1ex}{.}\hspace{.1em}\raisebox{-0.2ex}{.} }

\newcommand{\NF}{N_{\rm F}}

\newcommand{\hs}[1]{\hspace{#1 mm}}

\newcommand {\beq}{\begin{eqnarray}}
\newcommand {\eeq}{\end{eqnarray}}

\newcommand{\bpm}{\begin{pmatrix}}
\newcommand{\epm}{\end{pmatrix}}
\newcommand{\ba}{\left( \begin{array}}
\newcommand{\ea}{\end{array} \right)}
\newcommand{\beqnn}{\begin{eqnarray*}}
\newcommand{\eeqnn}{\end{eqnarray*}}

\newcommand{\R}{\mathbb{R}}
\newcommand{\C}{\mathbb{C}}

\newcommand{\diag}{{\rm diag}\,}

\newcommand{\baryon}{B^{n_1 n_2 \cdots n_k}_{\hspace{1pt} r_1 \, r_2 \hspace{1pt} \cdots \hspace{1pt} r_k}}

\makeatletter
\@addtoreset{equation}{section}

\renewcommand{\thefootnote}{\fnsymbol{footnote}}
\makeatother

\makeatletter
\newcommand{\thetablename}{Table}
\def\fnum@table{\thetablename\ \thetable}
\makeatother

\begin{document}
\begin{titlepage}
\begin{flushright}
IFUP-TH/2010-25\\
KUNS-2299\\
RIKEN-MP-6\\

September, 2010
\end{flushright}
\begin{center}
{\LARGE Group Theory of Non-Abelian Vortices}

\vspace{0.5cm}
Minoru Eto${}^{1}$, 
Toshiaki Fujimori${}^{2,3,4}$, 
Sven Bjarke Gudnason${}^{4,3}$,  \\
Yunguo Jiang${}^{4,3}$, 
Kenichi Konishi${}^{4,3}$,  
Muneto Nitta${}^{5}$, 
Keisuke Ohashi${}^{6}$

\vspace{0.5cm}
{\it\small 
${}^1$ Mathematical Physics Laboratory, Nishina Center, 
RIKEN, Saitama 351-0198, Japan}\\
{\it\small 
${}^2$ Department of Physics, Tokyo Institute of
Technology, Tokyo 152-8551, Japan}\\
{\it\small 
${}^3$ INFN, Sezione di Pisa, 
Largo B.~Pontecorvo, 3,  Ed.~C, 56127 Pisa, Italy}\\
{\it\small 
${}^4$ Department of Physics,``E. Fermi'', University of Pisa, 
Largo B.~Pontecorvo, 3, Ed.~C, 56127 Pisa, Italy}\\
{\it\small
${}^5$ Department of Physics, and Research and Education 
Center for Natural Sciences,}\\  
{\it\small 
 Keio University, Hiyoshi 4-1-1, Yokohama, Kanagawa 223-8521, Japan}\\
${}^6$ {\it\small 
Department of Physics, Kyoto University, Kyoto 
606-8502, Japan}

\vspace{1cm}
{\bf Abstract}

\vspace{0.5cm}
\parbox{15cm}{
\small\hspace{15pt}
We investigate the structure of the moduli space 
of multiple BPS non-Abelian vortices 
in $U(N)$ gauge theory with $N$ fundamental Higgs fields, 
focusing our attention on the action of 
the exact global (color-flavor diagonal) $SU(N)$ symmetry on it. 
The moduli space of a single non-Abelian vortex, $\C P^{N-1}$,  
is spanned by a vector in the fundamental representation of 
the global $SU(N)$ symmetry. 
The moduli space of winding-number $k$ vortices is 
instead spanned by vectors in the direct-product representation:  
they decompose into the sum of irreducible representations 
each of which is associated with a Young tableau made of $k$ boxes, in a way somewhat similar to
 the standard group composition rule of $SU(N)$ multiplets. 
The K\"ahler potential is exactly determined in each moduli subspace,
corresponding to an irreducible $SU(N)$ orbit of the highest-weight configuration. 
}
\end{center}
\end{titlepage}

\setcounter{footnote}{0}
\renewcommand{\thefootnote}{\arabic{footnote}}


\tableofcontents

\section{Introduction and discussion \label{sec:Intro}}

 Non-Abelian vortices have been discovered 
several years ago in the context of $U(N)$ supersymmetric gauge theories 
and in string theory \cite{Hanany:2003hp,Auzzi:2003fs}. 
BPS non-Abelian vortices exist in the $U(N)$ Yang-Mills
theory coupled to $\NF=N$ Higgs fields in the fundamental representation. 
The BPS equations are of the form
\beq
(\D_1 \pm i \D_2) H = 0, \qquad
F_{12} = \pm \frac{g^2}{2}\left( HH^\dagger - v^2{\bf 1}_N \right),
\label{eq:bps}
\eeq
where the upper (lower) sign describes the vortices (anti-vortices).   
The Higgs fields $H$ are combined in a color-flavor mixed $N\times N$ 
matrix on which the $U(N)$ gauge (color) symmetry acts on the left  
while the $SU(N)$ flavor symmetry acts on the right. 
The constant $g$ is the $U(N)$ gauge coupling\footnote{
Here we take common gauge couplings for $SU(N)$ and $U(1)$ of
$U(N)=[U(1)\times SU(N)]/\mathbb Z_N$ for simplicity.} 
and $v^2$ is the Fayet-Iliopoulos parameter. 
The $U(N)_{\rm C}$ gauge (color) symmetry is spontaneously broken
completely in the so-called color-flavor-locked vacuum
($\langle H\rangle = v{\bf 1}_N$), 
whereas  the global diagonal symmetry $SU(N)_{\rm C+F}$ remains unbroken. 
Since winding-number $k$ vortices 
(``$k$ vortices''  from now on, for simplicity) saturate the BPS energy (tension) bound
\beq
T  \ge 2 \pi v^2 k,
\eeq
no net forces are exerted among the static vortices. 
This implies that a set of solutions to Eq.\,\eqref{eq:bps} contains
integration constants, i.e.~moduli parameters parametrizing the set of
configurations with degenerate energy, viz.~the moduli space of BPS
vortices: $\mathcal M_k$. 

In addition to the position moduli, each non-Abelian vortex has
internal orientational moduli which are associated with the
$SU(N)_{\rm C+F}$ color-flavor symmetry, 
broken by the individual vortex configurations. 
Consider for instance a particular BPS solution
\beq
H = \diag\left( H^{\rm ANO} , v , \cdots , v \right), \qquad
A_\mu = \diag \left( A_\mu^{\rm ANO} , 0 , \cdots ,0 \right),
 \label{eq:particular}
\eeq
where $H^{\rm ANO}$ and $A_\mu^{\rm ANO}$ are the fields describing
the well-known Abrikosov-Nielsen-Olesen (ANO) vortex solution.  
Clearly, the solution breaks $SU(N)_{\rm C+F}$ down to 
$SU(N-1) \times U(1)$ 
and therefore the corresponding Nambu-Goldstone zero-modes, 
which we call internal orientational modes, 
appear on the vortex and parametrize the coset 
\beq 
\frac{SU(N)}{SU(N-1) \times U(1)} \cong \mathbb{C}P^{N-1}, 
\eeq
whose size (K\"ahler class) is given by $4\pi/g^2$
\cite{Hanany:2004ea,Shifman:2004dr,Eto:2004rz,Gorsky:2004ad}.  
The generic vortex solutions can be obtained 
by acting on the above solution with $U \in SU(N)_{\rm C+F}$, 
i.e., $H \rightarrow U^\dagger H U,~A_\mu \rightarrow U^\dagger A_\mu U$.
We parametrize them by using a normalized complex $N$-vector 
$\vec{\phi}\,~(\vec{\phi}^\dagger \cdot \vec{\phi} = 1)$ 
\beq
H = v \mathbf 1_N + (H^{\rm ANO} - v) \vec{\phi} \vec{\phi}^\dagger, \hs{10}
A_\mu = A_\mu^{\rm ANO} \vec{\phi} \vec{\phi}^\dagger. 
\eeq
Since the overall $U(1)$ phase of $\vec{\phi}$ is unphysical, 
the vector $\vec{\phi}$ can indeed be interpreted as 
the homogeneous coordinates of $\mathbb{C}P^{N-1}$.  
The moduli space of multiple vortices was found in
Ref.~\cite{Eto:2005yh} 
in terms of the moduli matrix formalism \cite{Isozumi:2004vg,Eto:2006pg}. 
The moduli space metric has been found recently for well-separated
vortices \cite{Fujimori:2010fk,Baptista:2010rv} 
using a generic formula for the K\"ahler potential 
on the moduli space \cite{Eto:2006uw}.

The starting point of our analysis is 
the observation that the vector $\vec{\phi}$ for a single vortex transforms 
according to the fundamental representation of $SU(N)_{\rm C+F}$. 
In this precise sense 
{\it the non-Abelian vortex belongs to 
the fundamental representation of $SU(N)_{\rm C+F}$.}  
Now the following question naturally arises:
{\it How does the color-flavor symmetry act in the moduli space $\mathcal M_k$, and to which representations do $k$ vortices belong?} 
Since each vortex has an orientational vector $\vec{\phi}_I~(I=1,\ldots,k)$ 
in the fundamental representation of $SU(N)_{\rm C+F}$, 
one expects that it is simply described by 
the tensor product of fundamental representations, e.g.,  
\beq 
\Young[0]{1} \otimes \Young[0]{1} = \Young[0]{2} \oplus \Young[-0.5]{11} ~.
\label{eq:product}
\eeq 
However, the situation is not so simple 
since the orientational vectors $\vec{\phi}_I$ are well-defined 
only when all vortices are separated. 
What happens when two or more vortices sit on top of each other? 
To answer these questions we must study the moduli space 
in such a way that allows a smooth limiting case 
where the vortex centers are taken to be coincident.  

The problem was already studied in the literature 
for $k=2$ coincident vortices in $U(2)$ gauge theory
\cite{Hashimoto:2005hi,Auzzi:2005gr,Eto:2006cx,Eto:2006db,Eto:2006dx,Auzzi:2010jt}, and partial answers were obtained.
While each vortex carries an orientation in ${\mathbb C}P^1$, 
the moduli space of two coincident vortices was found to be 
$W{\mathbb C}P^2_{(2,1,1)}\simeq {\mathbb C}P^2/{\mathbb Z}_2$ 
\cite{Hashimoto:2005hi,Auzzi:2005gr,Eto:2006cx}. 
In this case, each vortex belongs to ${\bf 2}$ 
so that the composition rule \eqref{eq:product} yields 
${\bf 2} \otimes {\bf 2} = {\bf 3} \oplus {\bf 1}$.
We have indeed found the moduli parameters transforming as ${\bf 3}$
and a singlet configuration corresponding to a ${\mathbb Z}_2$
singularity \cite{Eto:2006cx}. 
However, the precise knowledge about the correspondence 
between the representations and points in the moduli space was lacking. 
In other words, we did not know the true meaning of 
the composition-decomposition Eq.\,\eqref{eq:product} at that time. 

These questions are clarified in the present paper.

These issues are actually intimately related to 
the question of the non-Abelian {\it monopoles}. 
Indeed, a $U(N)$ vortex system such as ours can always be regarded as 
a low-energy approximation of an underlying larger, 
e.g. $SU(N+1)$, gauge theory, spontaneously broken to 
$SU(N) \times U(1)$ gauge group, 
by the vacuum expectation value (VEV) of 
some other scalar field at a mass scale much higher than 
the typical vortex mass scale. 
In such a hierarchical symmetry-breaking setting, 
whatever properties we find out about the vortices can be translated into 
those of the massive monopoles sitting at the extremes, 
as a homotopy-sequence consideration relates the two, 
at least semi-classically \cite{Eto:2006dx}\footnote{The monopole-vortex correspondence becomes far subtler when one is interested in the properties of light monopoles. The low-energy dynamics and renormalization-group effects both for the vortex \cite{Hanany:2004ea,Shifman:2004dr} and monopole \cite{SW} must be properly taken into account. This requires also a careful identification of the quantum vacua \cite{DKO}, as many of the systems involved possess large {\it vacuum moduli}. }. We shall, however, not dwell much on these points in the present work: we shall come back to them elsewhere.  

In this paper the moduli space of $k$ vortices are studied 
by using the $U(k)$ K\"ahler quotient construction due to 
Hanany-Tong \cite{Hanany:2003hp}. 
We analyze the moduli space in algebraic geometry 
by using certain $SL(k,{\mathbb C})$ invariants: 
symmetric polynomials of the vortex centers and 
``baryonic invariants'' \footnote{Although they have nothing to do with real physical baryons, for formal similarity and for convenience these invariants will be referred to  ``baryonic invariants'' or simply as ``baryons'': see Section~\ref{sec:invariants} below.}. 
We find algebraic constraints for these invariants 
which specify the embedding of the internal moduli space 
in a complex projective space. 
The moduli space of vortices contains various $SU(N)$ orbits, 
each of which belongs to a certain representation of $SU(N)$.  
We analyze the structures of those $SU(N)$ orbits by using 
``vortex state vectors'' constructed from the $SL(k,\C)$ invariants, 
by the help of some auxiliary harmonic-oscillator states. 

When $k$ vortices are all separated, 
vortex states can be written as coherent states in such a description. 
Accordingly, the vortex states can be shown to correspond to factorized 
(non-entangled) products of $k$ single vortex states 
in the fundamental representation. 

The situation of the $k$-winding vortices with coincident centers 
turns out to be considerably subtler. 
It will be shown that each $SU(N)$ orbit of 
$k$ rotationally invariant (axially symmetric) vortices
corresponding to some irreducible representation, 
which we call the ``irreducible $SU(N)$ orbit,'' 
can be classified by a Young tableau with $k$ boxes. 
Generic orbits belong to reducible representations and the associated 
vortex states can be written as a superposition of irreducible states.

One of the deepest aspects of our results is 
the fact that the vortex moduli, 
which describe a degenerate set of classical extended field configurations, 
behave under the exact $SU(N)$ global symmetry 
as a moduli space of quantum oscillator states, 
characterized by irreducible multiplets and 
having the possibility of superposition of ``states''.  
Even if this should be regarded just as a formal aspect of 
mathematical interest here, 
it could provide a physical key to quantum-mechanical understanding of 
non-Abelian monopoles through the vortex-monopole connection, 
briefly mentioned above. 

Also, albeit our results here  -- understandably --  basically obey
the standard composition rule for $SU(N)$ multiplets, 
the composition rule of the non-Abelian vortices  is found to possess  various special features
(see below);  for instance, the vortex moduli space involves in general much fewer dimensions 
than na\"ively expected. 

All irreducible $SU(N)$ orbits are K\"ahler submanifolds 
inside the full moduli space. 
We shall construct the K\"ahler potential 
on each of the irreducible $SU(N)$ orbits and 
find that the coefficient of the K\"ahler potential is
quantized as an integer: 
the latter is uniquely specified by the associated Young tableau. 
We point out the existence of a duality between pairs of irreducible orbits corresponding to the conjugate representations of $SU(N)$, 
which are found to describe, as expected, the same low-energy effective action.

The rest of the paper is organized as follows. 
In Section\,\ref{sec:msnavor}, 
the basic features of the moduli space of $k$
non-Abelian vortices are reviewed. 
We then proceed to construct the ``baryonic invariants'' 
which form good coordinates on our moduli space. 
By making use of these we  find
the representations of $k$ separated vortices in Section\,\ref{sec:sepa};  
we construct an irreducible representation for 
a specific (highest-weight) configuration of coincident vortices 
in Section\,\ref{sec:sunorbitscoincidentvor}.  In 
Section\,\ref{sec:examples} the solution to the constraints on the ``baryonic invariants'' is worked out and the result is used to show  
the $SU(N)$ decomposition rule for generic vortex solutions for given $k$.   A particular attention is paid to the consideration of the limit of 
co-axial vortices.   The cases of  
$k=1,2,3$ are explicitly solved, while a general recipe is given,  valid for any $N, k$.  
The K\"ahler potentials for the irreducible $SU(N)$ orbits are
obtained in Section\,\ref{sec:kahler}. 
A brief summary and outlook is given in Section\,\ref{sec:conclusion}. 
A few details of our analysis are postponed to the Appendices.

\section{Moduli space of non-Abelian vortices}
\label{sec:msnavor}

\subsection{The moduli space and $GL(k,\mathbb{C})$
  invariants \label{sec:invariants}  } 

The moduli space of the non-Abelian $U(N)$ vortices governed by 
the BPS Eq.\,\eqref{eq:bps} was first studied by 
Hanany-Tong \cite{Hanany:2003hp}. 
There the dimension of the moduli space ${\cal M}_k$ of $k$ vortices 
has been shown by using an index theorem calculation to
be\footnote{The general result of Ref.~\cite{Hanany:2003hp} in $U(N)$
  theory for $\NF\geq N$ flavors is 
  ${\rm dim}_{\mathbb{C}} \mathcal{M}_k = k \NF$. However, we restrict
  our attention to the case $\NF=N$ and hence local vortices in this
  paper. }  
\beq
{\rm dim}_{\mathbb{C}} \, {\cal M}_k = k N,
\eeq
with $k$ being the topological winding number.
Moreover, they found a D-brane configuration and 
derived a K\"ahler quotient construction for ${\cal M}_k$.  
It is sometimes called {\it a half-ADHM construction} 
by analogy with the moduli space of instantons. 
In the D-brane configuration, the $k$ vortices are 
$k$ D2-branes suspended between $N$ D4-branes and an NS5-brane. 
The low-energy effective field theory on the $k$ D2-branes is
described by a $U(k)$ gauge theory 
coupled with a $k$-by-$k$ matrix $Z$ in adjoint representation 
and a $k$-by-$N$ matrix $\psi$ in the fundamental representation 
${\bf k}$ of the $U(k)$ gauge symmetry,
given by D2--D2 strings and D2--D4 strings, respectively.
The $U(k)$ gauge symmetry on the D2-branes acts on $Z$ and $\psi$ as 
\beq
(Z,\psi) \to \left(g Z g^{-1} , g \psi \right), \qquad g \in U(k).
\label{eq:u(k)}
\eeq
The moduli space ${\cal M}_k$ can be read off 
as the Higgs branch of vacua in the $U(k)$ gauge theory on the $k$ D2-branes, 
which is the K\"ahler quotient of the $U(k)$ action \eqref{eq:u(k)}\footnote{
Here the normalization of the scalar fields $Z,\,\psi$ is chosen so
that they have canonical kinetic terms in the 2-dimensional
effective gauge theory on the D2 branes. In this convention the
eigenvalues of $Z$ (i.e.~vortex positions) are dimensionless
parameters.}  
\beq
{\cal M}_k ~\,\cong\,~ 
{\cal M}_k^{\rm HT}  &\equiv& 
\left\{ \, (Z,\psi) \ | \ \mu_D = r{\bf 1}_k \,\right\} /
U(k),\label{eq:HTMdu} \\ 
\mu_D ~&\equiv& [ Z , Z^\dagger ] + \psi \psi^\dagger.
\label{eq:HT}
\eeq
This K\"ahler quotient gives a natural metric on ${\cal M}_k$ 
provided that $(Z,\psi)$ has a flat metric on $\mathbb{C}^{k(k+N)}$.
Unfortunately, the geodesics of such a metric do not describe 
the correct dynamics of vortices \cite{Hanany:2003hp}. 
The 2d FI parameter $r$ is related to the 4d gauge coupling constant by
\beq
r = \frac{4\pi}{g^2},
\label{eq:kahlerclass}
\eeq
which holds under the RG flow 
if the 4d theory has ${\cal N}=2$ supersymmetry 
and the 2d theory has ${\cal N}=(2,2)$ supersymmetry 
\cite{Hanany:2004ea,Shifman:2004dr}. 

According to Ref.~\cite{Hitchin:1986ea} 
the K\"ahler quotient \eqref{eq:HTMdu} can be rewritten 
as a complex symplectic quotient as 
\beq
{\cal M}_k &\cong& \left\{ \, (Z,\psi) \, \right\}
/\!\!/GL(k,\mathbb{C}),
\label{eq:kq}
\eeq
where instead of having the $D$-term condition $\mu_D = r \mathbf 1_k$, 
the pair of matrices $(Z,\psi)$ is divided by 
the complexified non-compact group $U(k)^{\mathbb{C}} = GL(k,\mathbb{C})$ 
which acts in the same way as Eq.\,\eqref{eq:u(k)}. 
Here the quotient denoted by the double slash ``$/\!\!/$'' means 
that points at which the $GL(k,\mathbb{C})$ action is not free 
should be removed so that the group action is free at any point. 
This quotient is also understood as the algebro-geometric quotient,
so that the quotient space is parametrized by a set of
$GL(k,\mathbb{C})$ holomorphic invariants with suitable constraints,
see e.g.~Ref.~\cite{Luty:1995sd}.

The starting point of our analysis, Eq.~(\ref{eq:kq}),  can also be obtained directly from a purely field-theoretic point of view,  
based on the BPS equation \eqref{eq:bps}.
It has been shown by using the moduli-matrix approach
\cite{Isozumi:2004vg,Eto:2005yh,Eto:2006pg} that 
all the moduli parameters of the $k$-vortex solutions  are  
summarized exactly  as in Eq.\,\eqref{eq:kq}.
The 4d field theory also provides the correct metric on ${\cal M}_k$
describing the dynamics of vortices 
as a geodesic motion on the moduli space. 
Although a general formula for the metric and its K\"ahler potential
has been derived \cite{Eto:2006uw}, 
the explicit form of the metric is however difficult 
to obtain since no analytic solutions to the BPS equation are known. 
Nevertheless, the asymptotic metric for well-separated vortices has
recently been found in Ref.~\cite{Fujimori:2010fk}.\footnote{ 
See Ref.~\cite{Baptista:2010rv} for an alternative formula for
vortices on Riemann surfaces.}

In what follows, we analyze the moduli space 
Eq.\,\eqref{eq:kq} without assuming any metric {\it a priori}.
Our prime concern is how the exact global $SU(N)$ symmetry acts 
on the vortex moduli space $\mathcal M_k$. 
The matrix $Z$ is a singlet while $\psi$ belongs to 
the fundamental representation $\mathbf N$. 
Namely, the $SU(N)$ acts on $Z$ and $\psi$ as
\beq
Z \to Z, \quad \psi \to \psi \, \mathcal U, \qquad \mathcal U \in SU(N).
\label{eq:u(n)}\eeq
As will be seen this action induces a natural $SU(N)$ action 
on the moduli space of vortices.  
We will also discuss the metrics on the symmetry orbits 
on which the $SU(N)$ acts isometrically. 
To this end, we use the algebro-geometric construction
\cite{Luty:1995sd} of the moduli space 
by using the $GL(k,\mathbb{C})$ invariants 
which provide a set of coordinates of the moduli space. 

Clearly, the coefficients $\sigma_i~(i=1,\ldots,k)$ of the
characteristic polynomial of $Z$ are invariants of the $GL(k,\C)$ action 
\beq
\det \left(\lambda {\bf 1}_k - Z \right) 
= \lambda^k + \sum_{i=1}^k (-1)^i \sigma_i \lambda^{k-i}. 
\label{eq:posi}
\eeq
Since the vortex positions $z_I~(I=1,\ldots,k)$ are defined as the
eigenvalues of $Z$ (roots of the characteristic polynomial)
\beq
\det \left(\lambda {\bf 1}_k - Z \right) = \prod_{I=1}^k (\lambda - z_I),
\eeq
the parameters $\sigma_i$ and $z_I$ are related by 
\beq
\sigma_i = P_i(z_1,\cdots,z_k),
\eeq
where $P_i~(i=1,\ldots,k)$ are the elementary symmetric polynomials
defined by 
\beq
P_i(z_1,\cdots,z_k) \equiv 
\sum_{1 \leq I_1 < \cdots < I_i \leq k} z_{I_1} z_{I_2} \cdots z_{I_i}.  \label{defelsym}
\eeq
Note that vortex positions $z_I$ are not fully invariant 
under $GL(k,\C)$ transformations since they can be exchanged 
by the Weyl group $\mathfrak{S}_k$. 

Other invariants can be constructed as follows. 
Let $Q^{(n)}~(n=0,1,\ldots)$ be the following 
$({\bf k},{\bf N})$ matrices   of $SL(k,\C) \times SU(N)$  (Eqs.~(\ref{eq:u(k)}) and (\ref{eq:u(n)})): 
\beq
Q^{(0)} \equiv \psi, \quad 
Q^{(1)} \equiv Z \psi, \quad 
\cdots , \quad 
Q^{(n)} \equiv Z^{n} \psi, \quad
\cdots.
\eeq
One can construct $SL(k,\mathbb{C}) \subset GL(k,\mathbb{C})$ 
invariants from $Q^{(n)}$ by using the totally anti-symmetric tensor
$\epsilon^{i_1 \cdots i_k}$ as 
\beq
\baryon \equiv \epsilon^{i_1 i_2 \cdots i_k} Q^{(n_1)}_{i_1 r_1}
Q^{(n_2)}_{i_2 r_2} \cdots Q^{(n_k)}_{i_k r_k}\;.
\label{eq:def_B}
\eeq
We call these the ``baryonic invariants'' 
or sometimes simply ``the baryons''  below, 
relying on a certain analogy to the baryon states 
in the quark model (or in quantum chromodynamics). 
 
\noindent{\bf Remark:} although obviously 
they have no physical relation to the real-world baryons 
(the proton, neutron, etc.), 
no attentive reader should be led astray by such a short-hand notation. 

Note that the baryons (\ref{eq:def_B}) are invariant under $SL(k,\C)$ 
and transform under the remaining $U(1)^{\C} \cong \C^\ast$ as 
\beq
\baryon ~\rightarrow~ e^{\lambda} \baryon ,
\label{eq:U1equiv}
\eeq
with a suitable weight $\lambda$. 

The vortex positions 
$\{z_I\} \cong {\mathbb C}^k/\mathfrak{S}_k \cong {\mathbb C}^k$ 
are parametrized by the moduli parameters 
$\{\sigma_i\} \cong {\mathbb C}^k$. 
In addition to these parameters, there are baryons 
\[ \{\baryon\} \cong V ,\] as moduli parameters, 
where $V$ denotes an infinite-dimensional complex linear space 
spanned by the baryons. 
The problem is that not all of these invariants are 
independent of each other; the baryons $\baryon$ and $\sigma_i$ satisfy 
certain constraints by construction. 
Therefore, the vortex moduli space Eq.\,\eqref{eq:kq} 
can be rewritten as 
\beq
\mathcal M_k ~\cong~ 
\{ \, \C^k \times V ~|~ \mbox{constraints} \ \} \, /\!\!/ \, \C^\ast.
\eeq
Since the baryonic invariants transform under $SU(N)$, 
there exists a linear action of $SU(N)$ on $V$: 
this induces an $SU(N)$ action on the moduli space.  

Consider now the constraints 
on the parameters $\sigma_i$ and the baryons $\baryon$ in more detail. 
For this purpose it turns out to be convenient to 
introduce an auxiliary set of $k$ linear harmonic oscillator states, 
each of which carrying an $SU(N)$ label, 
and make a map from the vector space $V$ to the Fock space of such
oscillators. Let us introduce a ``vortex state vector''\footnote{We
  hasten to add that no relation between the notion of vortex ``state
  vectors'' here and any quantum dynamics is implied by such a
  construction. } 
$|B\rangle \in V$ by
\beq
\left| B \right> ~\equiv 
\sum_{{}^{n_1,}_{\hspace{1pt}r_1,}{}^{n_2,}_{\hspace{1pt}r_2,}{}^{\cdots,}_{\cdots,}{}^{n_k}_{\hspace{1pt}r_k}} 
\frac{1}{(n_1! n_2!\cdots n_k!)^{\frac{1}{2}}} 
\baryon \left| n_1,r_1 \right> \otimes 
\left|n_2, r_2 \right> \otimes \cdots \otimes 
\left| n_k, r_k \right>,
\eeq
with $n_i\in {\mathbb Z_{\ge 0}}$,~$1 \le r_i \le N$;  
the associated annihilation and creation operators 
$\hat a_i,\hat a_i^\dagger~(i=1,\ldots,k)$ 
\beq
\hat a_i \Big( \cdots \otimes \left| n_i,r_i \right> \otimes 
\cdots \Big) 
&=& \hs{4} \sqrt{n_i} \hs{4} \Big( \cdots \otimes 
\left| n_i-1,r_i \right> \otimes \cdots \Big), \\
\hat a_i^\dagger \Big( \cdots \otimes \left| n_i,r_i \right> \otimes 
\cdots \Big) 
&=& \sqrt{n_i+1} \hspace{2pt} \Big( \cdots \otimes 
\left| n_i+1,r_i \right> \otimes \cdots \Big)\; 
\eeq
satisfy the standard commutation relations 
\beq
[\hat a_i, \hat a_j^\dagger ] = 
\delta_{ij}, \hs{10} [\hat a_i, \hat a_j ] = 
[ \hat a_i^\dagger, \hat a_j^\dagger ] = 0\;.
\eeq
Note that once $\left| B \right>$ is given, 
the baryonic invariants can be read off from the following relation
\beq
\baryon = \left< 0,r_1 \, ; \cdots ; 0,r_k \right| (\hat a_1)^{n_1}
(\hat a_2)^{n_2} \cdots (\hat a_k)^{n_k} \left| B \right>, 
\label{eq:state_to_B}
\eeq
where $\left| 0,r_1 \, ; \cdots ; 0,r_k \right> \equiv 
\left|0,r_1 \right> \otimes \cdots \otimes \left|0,r_k \right>$ 
are the ground states.
Now there are three types of constraints to be taken into account 
(see Appendix \ref{appendix:constraints} for more details):  
\begin{enumerate}
\item
From definition \eqref{eq:def_B} one can see
that the baryons satisfy the anti-symmetry property
\beq
B^{A_1 \cdots A_i \cdots A_j \cdots A_k} = -B^{A_1 \cdots A_j \cdots A_i \cdots A_k},
\label{eq:anti-symmetric}
\eeq
where $A_i$ stands for the pair of indices $(n_i,r_i)$. 
This constraint can be rewritten as
\beq
\hat \rho \, \left| B \right> = {\rm sign}(\rho) \left| B \right>,
\label{eq:constraint1}
\eeq
where $\hat{\rho}$ denotes an element of the symmetric group $\mathfrak{S}_k$. 
For an element $\hat{\rho} \in \mathfrak{S}_k$ 
\beq
\rho = \ba{cccc} 1 & 2 & \cdots & k \\ I_1 & I_2 & \cdots & I_k \ea,
\eeq
the action on the state is defined by
\beq
\hat{\rho} \left|n_1,r_1 \right> \otimes 
\left|n_2,r_2 \right> \otimes \cdots \otimes 
\left|n_k,r_k \right> ~=~ 
\left|n_{I_1},r_{I_1} \right> \otimes 
\left|n_{I_2},r_{I_2} \right> \otimes \cdots \otimes 
\left|n_{I_k},r_{I_k} \right>.
\eeq
\item
The second condition is a consequence of  the relation 
$Q^{(n+m)}=Z^m Q^{(n)}$.  
It follows that
\beq
P_i(\hat a_1,\cdots,\hat a_k) \left| B \right> = 
\sigma_i \left| B \right>, \qquad (i=1,\ldots,k),
\label{eq:constraint2}
\eeq
where $P_i(\hat a_1,\cdots,\hat a_k)$ are 
the elementary symmetric polynomials made of $\hat a_i$  (cfr. Eq.~(\ref{defelsym})).

\item
The last type of constraints are the quadratic equations 
for the baryons, which follow from Eq.\,\eqref{eq:def_B}: 
\beq
B^{A_1 A_2 \cdots A_{k-1}[A_k} B^{B_1B_2\cdots B_k]} = 0,
\label{eq:constraint3}
\eeq
where $A_i$ stands for the pair of indices $(n_i,r_i)$. 
This constraint is a generalization of the Pl\"ucker relations 
for the Grassmannian.  
\end{enumerate}
Eqs.\,\eqref{eq:constraint1} and \eqref{eq:constraint2} can be viewed
as linear constraints for baryons with $\sigma_i$-dependent coefficients.  
Therefore, for a given set of values $\{\sigma_i\}$, 
they define a linear subspace $W(\sigma_i) \subset V$ 
to which the vortex state vector $\left|B \right>$ belongs. 
We will see that the representation of the $SU(N)$ action on
$W(\sigma_i)$ is independent of $\sigma_i$ and isomorphic to $k$
copies of the fundamental representation $\mathbf{N}$ 
\beq
W(\sigma_i) \,~\cong~\, {\mathbb C}^{N^k} ~\cong~
\bigotimes_{i=1}^k \mathbf N.
\label{eq:direct_product}
\eeq
Note that not all vectors in this ``state space" $W(\sigma_i)$ represent  vortex state vectors 
since they must still satisfy Eq.\,\eqref{eq:constraint3}.
Namely, the vortex moduli space is defined by the constraints
\eqref{eq:constraint3}, which are quadratic homogeneous polynomials of
the coordinates of $W( \sigma_i )$ with $\sigma_i$-dependent coefficients.

\subsection{The moduli space of $k$ separated vortices}
\label{sec:sepa}

Let us first consider the case of winding-number $k$ vortices 
with distinct centers, $z_I \neq z_J$ (for all $I \neq J$). 
It follows from Eq.\,\eqref{eq:constraint2} that for $i=1,2,\ldots,k$ 
\beq
\prod_{I=1}^k (\hat{a}_i - z_I) \left| B \right> = 
\left((\hat{a}_i)^k+\sum_{n=1}^{k}(-1)^n \sigma_n (\hat{a}_i)^{k-n}
\right) 
\left| B \right> = \prod_{j=1}^k(\hat{a}_i-\hat{a}_j)\left|B \right>=0.
\label{eq:constraint4}
\eeq
Thus, in the case of $z_I \neq z_J$,
there exists an $I_i~(1 \leq I_i \leq k)$ for each $i$ such
that\footnote{Note that this relation 
does not necessarily hold for coincident vortices. 
For example, if $z_I = z_J = z_0~(I \neq J)$, the constraint
\eqref{eq:constraint4} can also be satisfied 
by a state vector $\left| B \right>$ such that
\beq
(\hat{a}_i - z_0)^2 \left| B \right> = 0, \hs{5} \hat{a}_i 
\left| B \right> \not = z_0 \left| B \right>. \nonumber
\eeq
}  
\beq
\hat a_i \left| B \right> = z_{I_i} \left| B \right>.
\label{eq:eigenstate}
\eeq
Namely, the most generic form of the solution to the constraint
\eqref{eq:constraint4} is 
\beq
\left| B \right> ~= 
\sum_{{}^{I_1,}_{r_1,}{}^{I_2,}_{r_2,}{}^{\cdots,}_{\cdots,}{}^{I_k}_{r_k}} 
\tilde{B}^{I_1 I_2 \cdots I_k}_{r_1 r_2 \cdots r_k} 
\left| z_{I_1}, r_1 \right> \otimes 
\left| z_{I_2}, r_2 \right> \otimes \cdots \otimes 
\left| z_{I_k}, r_k \right>, 
\label{eq:coherent}
\eeq
where $\left| z_{I_i}, r_i \right>$ are the coherent states defined by 
\beq
\left| z_{I_i}, r_i \right> \equiv 
\exp\left( z_{I_i} \hat a_i^\dagger \right) \left| 0 , r_i \right>.
\eeq
Recall that the coherent states are eigenstates of the annihilation
operators 
\beq
\hat a_i \left| z_{I_i}, r_i \right> = 
z_{I_i} \left| z_{I_i}, r_i \right>.
\eeq
Then the constraint \eqref{eq:constraint2} reads
\beq
P_i(z_{I_1}, z_{I_2}, \cdots, z_{I_k}) \left| B \right> = 
\sigma_i \left| B \right> ~~\Big( = P_i(z_1, z_2, \cdots, z_k) 
\left| B \right> \Big).
\eeq
This means that $\{z_{I_1},z_{I_2},\cdots,z_{I_k}\}$ is a
permutation of $\{z_1,z_2,\cdots,z_k\}$.  
Taking into account the anti-symmetry condition \eqref{eq:constraint1}, 
the solution of the constraints 
\eqref{eq:constraint1} and \eqref{eq:constraint2} is given by 
\beq
\left| B \right> ~= 
\sum_{r_1,r_2,\cdots,r_k} \tilde B_{r_1 r_2 \cdots r_k} \, 
\hat{\mathcal{A}} \Big( \left| z_1, r_1 \right> \otimes 
\left| z_2, r_2 \right> \otimes \cdots \otimes 
\left| z_k, r_k \right> \Big),
\label{eq:def_tB}
\eeq
where $\hat{\mathcal{A}}$ denotes 
the anti-symmetrization of the states  
\beq
\mathcal{\hat A} ~\equiv~ 
\frac{1}{k!} \sum_{\rho \in \mathfrak{S}_k}
{\rm sign}(\rho) \, \hat{\rho}.
\eeq 
For a given set $\{z_1,z_2,\cdots,z_k\}$, 
the solutions \eqref{eq:def_tB} span 
an $N^k$-dimensional vector space $W(\sigma_i)$ and 
the redefined baryons $\tilde{B}_{r_1 r_2 \cdots r_k}$ are 
the coordinates of $W(\sigma_i)$.  
As stated in Eq.\,\eqref{eq:direct_product}, 
$\tilde B_{r_1 r_2 \cdots r_k}$ is in the direct product
representation $\bigotimes_{i=1}^k \mathbf{N}$. 
They can be expressed in terms of 
the original baryons $\baryon$ by using the relation 
\beq
\tilde B_{r_1 r_2 \cdots r_k} = 
\left< 0, r_1 \, ; \cdots ; 0, r_k \right| 
e_1(\hat a_1) \cdots e_k(\hat a_k) \left| B \right>, 
\eeq
where $\left| 0,r_1 \, ;\cdots ; 0,r_k \right> \equiv 
\left|0,r_1 \right> \otimes \cdots \otimes \left|0,r_k \right>$ are
the ground states and $e_I~(I=1,\ldots,k)$ are 
the polynomials defined as
\beq
e_I(\lambda) \equiv \prod_{J \neq I} 
\frac{\lambda-z_J}{z_I - z_J},  \quad 
\left(e_I(z_J)=\delta_{IJ}\right).
\eeq
Since this polynomial is ill-defined for 
coincident vortices $z_I=z_J$ (for $I\not=J$), 
the coherent state representation \eqref{eq:def_tB} is 
valid only for separated vortices. 
As we will see later, 
there exist well-defined coordinates of $W(\sigma_i)$ 
for arbitrary values of $\sigma_i$.  
They can be obtained from $\tilde B_{r_1 r_2 \cdots r_k}$ 
by linear coordinate transformations with $z_I$-dependent coefficients. 
Hence the result that the linear space $W(\sigma_i)$ has 
the representation $\bigotimes_{i=1}^k \mathbf{N}$ holds for 
arbitrary values of $\sigma_i$, 
including the coincident cases ($z_I = z_J$), as well.

So far we have specified the state space $W(\sigma_i)$ to which 
the vortex state vectors belong. 
Now let us examine which vectors in $W(\sigma_i)$ 
can be actually allowed as vortex state vectors. 
The remaining constraint is the Pl\"ucker relation
\eqref{eq:constraint3} which reads
\beq
\tilde B_{r_1 \cdots r_i \cdots r_k}
\tilde B_{s_1 \cdots s_i \cdots s_k}
=
\tilde B_{r_1 \cdots s_i \cdots r_k}
\tilde B_{s_1 \cdots r_i \cdots s_k},
\eeq
for each $i=1,2,\ldots,k$. This is solved by
\beq
\tilde B_{r_1 r_2 \cdots r_k} = \phi^1_{r_1}\phi^2_{r_2}\cdots \phi^k_{r_k},
\label{eq:baryon_sep}
\eeq
Since the baryons are divided by 
$U(1)^{\mathbb C} \subset GL(k,\mathbb{C})$, 
the multiplication of a non-zero complex constant on each 
of $\vec{\phi}^I \in \mathbb{C}^N~(I=1,\ldots,k)$ is unphysical.
Therefore, each $N$-vector $\vec{\phi}^I = (\phi^I_{1},\cdots ,\phi^I_N)$ 
parametrizes $\mathbb{C}P^{N-1}$. 

We thus see that for separated vortices 
the baryon given in Eq.\,\eqref{eq:def_tB} 
can be written as an anti-symmetric product of ``single vortex states'' 
\beq
\left| B \right> ~= \hat{\mathcal{A}} 
\left[ \left( \sum_{r_1=1}^N \phi^1_{r_1} 
\left| z_1, r_1 \right> \right) \otimes 
\left( \sum_{r_2=1}^N \phi^2_{r_2} 
\left|z_2, r_2 \right> \right) \otimes \cdots \otimes 
\left( \sum_{r_k=1}^N \phi^k_{r_k} 
\left|z_k, r_k \right> \right) \right].
\label{farless}   \eeq
This means that the moduli space of the separated vortices is 
just a $k$-symmetric product of $\C \times \C P^{N-1}$ parametrized
by the position of the vortices $z_I$ and the orientation $\vec\phi^I$
\cite{Eto:2005yh}
\beq
{\cal M}^{k\text{-separated}} \simeq 
\left(\mathbb{C}\times \mathbb{C}P^{N-1}\right)^k/\mathfrak{S}_k,   \label{dirprod}
\eeq
where $\mathfrak{S}_k$ stands for the symmetric group. 
Note that the space of vortex states Eq.~(\ref{farless}), 
which are just generic (anti-symmetrized) {\it factorized} states. It 
spans far fewer dimensions ($2 N k$) 
than might na\"ively be expected for the product-states made of $k$ vectors, 
which would have a dimension of the order of $ 2 N^{k}$, 
ignoring the position moduli.  

\subsection*{Remarks} As is clear -- hopefully -- from our construction,   the use of the vortex ``state vector'' notion  
is here for convenience only, made for exhibiting the group-theoretic properties of the non-Abelian vortices. In other words we do not attribute to $\left|B \right>$   any direct physical significance. Accordingly, we need not discuss the question of their normalization 
(metric on the vector space $V$) here.  Note that two of the constraints (Eq.~(\ref{eq:constraint1}) and Eq.~(\ref{eq:constraint2})) are indeed linear;   the third, quadratic
constraint   (Eq.~(\ref{eq:constraint3})) does not affect their normalization either. 

  It is tempting,  on the other hand, to note that  
any choice of a metric in $V$ would induce a metric on the vortex moduli space, which is of physical interest.  
As discussed briefly in Appendix \ref{sec:toy}, however, a simple-minded choice of the metric for  $\left|B \right>$ does not lead to the fully  
correct behavior of the vortex interactions.

\subsection{Highest-weight coincident vortices and $SU(N)$ irreducible orbits}
\label{sec:sunorbitscoincidentvor}

Let us next consider $k$ vortices on top of each other, 
all centered at the origin. 
Namely we focus our attention on the subspace of the moduli space 
specified by the condition 
\beq
\sigma_i=0 \hs{5} \mbox{for all $i$}.
\eeq
Since the coherent states of Eq.\,\eqref{eq:coherent} are 
not the general solution to the constraint \eqref{eq:constraint4}, 
the situation is now more complicated. 
To understand the structure of this subspace in detail, 
it is important to know how the $SU(N)_{C+F}$ acts on it.  
As we have seen, the moduli space of vortices can be described 
in terms of the vortex state vector endowed with 
a linear representation of the $SU(N)$ action. 
We will denote the $SU(N)$ orbits of highest-weight vectors 
(to be defined below) the {\it ``irreducible $SU(N)$ orbits''} 
since the vectors belong to irreducible representations on those orbits. 
In this subsection we classify irreducible $SU(N)$ orbits 
by Young tableaux. 

\begin{figure}[!t]
\begin{center}
\includegraphics[width=90mm]{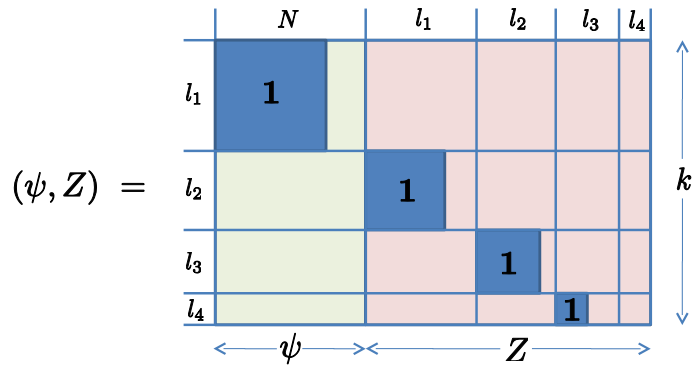}
\caption{{\small An example of a $k$-by-$(N+k)$ matrix $(\psi,Z)$ with
  $k_1=4$. The painted square boxes stand for unit matrices while the
  blank spaces imply that all their elements are zero. }} 
\label{fig:psi_Z}
\end{center}
\end{figure}

The ``highest-weight vectors''  will be defined 
as the special configurations of $\psi$ and $Z$ 
satisfying the following conditions:  
\begin{itemize}
\item 
Any $U(1)^{N-1}$ transformation in the Cartan subgroup of $SU(N)$ 
can be absorbed by a $GL(k,\C)$ transformation. 
Namely, for an arbitrary diagonal matrix $D \in U(1)^{N-1}$, 
there exists an element $g \in GL(k,\C)$ such that 
\beq
\psi \, D = g \, \psi, \qquad Z = g Z g^{-1}.
\label{eq:cond1}
\eeq
\item 
Any infinitesimal $SU(N)$ transformation 
with a raising operator $\hat E_\alpha$ can be 
absorbed by an infinitesimal $SL(k,\C)$ transformation. 
Namely, for an arbitrary lower triangular matrix $L$ 
whose diagonal entries are all 1, 
there exists an element $\tilde g \in SL(k,\C)$ such that 
\beq
\psi \, L = \tilde g \, \psi, \hs{10} Z = \tilde g Z \tilde g^{-1}.
\label{eq:cond2}
\eeq
\end{itemize}
Such configurations are classified by 
a non-increasing sequence of integers $\{l_1,l_2,\cdots,l_{k_1}\}$ satisfying
\beq
N \geq l_1 \geq l_2 \geq \cdots \geq l_{k_1} \geq 0, \qquad 
l_1 + l_2 + \cdots + l_{k_1} = k. 
\label{eq:highest}
\eeq
In other words, they are specified by Young tableaux (diagrams)\footnote{
In the following, the term ``Young tableaux" is used to denote diagrams without numbers in the boxes (Young diagrams), unless otherwise stated.} with $k$ boxes
\beq
\def\YGbox{20}
\overbrace{
\YoungTab[-2.5][\scriptsize]{{1,1,\cdots,\cdots,1} 
{2,2,\cdots,2} 
{{}^{\vdots},{}^{\vdots},} 
{{}^{\vdots},l_2} 
{l_1}}}^{k_1}
\label{eq:samp_yt}
\eeq
where the height of the $i$-th column is $l_i$ 
and the width of the $i$-th row is $k_i$. 
The total number of boxes is equal to 
the vortex winding number $k$. 
An example of a pair of matrices $(\psi,Z)$ 
corresponding the highest-weight state is given in Fig.\,\ref{fig:psi_Z}. 
For such a pair of matrices $(\psi,Z)$, 
one can check the existence of $g$ and $\tilde{g}$ satisfying 
Eq.\,\eqref{eq:cond1} and Eq.\,\eqref{eq:cond2}, given by 
\beq
g = 
\begin{pmatrix}
D_{l_1} & & \\ 
& \ddots & \\ 
& & D_{l_{k_1}} 
\end{pmatrix} , \qquad
\tilde{g} = 
\begin{pmatrix} 
L_{l_1} & & \\ 
& \ddots & \\ 
& & L_{l_{k_1}} 
\end{pmatrix}, 
\label{eq:def_g}
\eeq
where $k_1$ is the number of boxes in the first row of the Young
tableau, and $D_{l_i}$ and $L_{l_i}$ are the upper-left $l_i$-by-$l_i$ minor
matrices of $D$ and $L$, respectively.\footnote{For example, 
$D_{l_i} = \diag ( e^{i \theta_1}, \cdots, e^{i \theta_{l_i}} )$ for 
$D = \diag ( e^{i \theta_1} , \cdots, e^{i \theta_N} )$.}

The baryons corresponding to $(\psi,Z)$ are given by 
\beq
\left| B \right> ~=~
 \hat{\mathcal{A}} \Big[ \left| l_1 \right> \otimes 
\left| l_2 \right> \otimes \cdots \otimes 
\left| l_{k_1} \right>
\Big], \qquad
\left| l_{n+1} \right> ~\equiv~ 
\left|n ,1 \right> \otimes \left|n ,2 \right> \otimes \cdots \otimes 
\left|n,l_{n+1} \right>.
\label{eq:sample}
\eeq
We claim that this state is the highest-weight vector 
of the irreducible representation of $SU(N)$ specified by the Young tableau. 
This can be verified as follows. 
Since $(\psi,Z)$ satisfy the condition Eq.\,\eqref{eq:cond1}, 
the baryons transform under the $U(1)^{N-1}$ transformation according to 
\beq
\left| B \right> ~\rightarrow~ 
\det g \left| B \right> 
= \exp \left( i \sum_{i=1}^{l_1} k_i \theta_i \right) \left| B \right>, 
\qquad 
\sum_{i=1}^N \theta_i=0 , 
\eeq
where $k_i$ is the number of boxes in the $i$-th row of the Young
tableau. 
The weights of the $U(1)^{N-1}$ action can be 
read off in terms of $k_i$ as 
\beq
m_i = k_i - k_{i+1},
\eeq
where the integers 
$\left[ m_1,m_2,\cdots,m_{N-1} \right]$ are the Dynkin labels. 
On the other hand, since $(\psi,Z)$ satisfy the condition
Eq.\,\eqref{eq:cond2}, the $SL(k,\C)$ invariants $\baryon$ are
annihilated by the raising operators
\beq
\hat E_\alpha \left| B \right> = 0.
\eeq
We have thus proved that
\eqref{eq:sample} represents the highest-weight state 
of the representation \eqref{eq:samp_yt} in the usual sense.

We define an ``{\it irreducible $SU(N)$ orbit for the set of
  Dynkin labels: $[m_1,m_2,\cdots,m_{N-1}]$ }'' 
as an $SU(N)$ orbit of the corresponding highest-weight state. 
Note that this definition is obviously independent of 
the choice of $U(1)^{N-1}\in SU(N)$ in Eq.\,\eqref{eq:cond1}.
It is known that such an orbit is a generalized flag manifold 
of the form $SU(N)/H$ with $H$ being a subgroup of $SU(N)$ 
which acts on the highest-weight state as 
\beq
\hat{h} \left| B \right> ~=~ 
e^{i \theta(\hat h)} \left| B \right> \sim \left| B \right>, \qquad
\forall \, \hat{h} \in H.
\eeq
The subgroup $H$ can be specified by removing the nodes 
in the Dynkin diagram which correspond to non-zero Dynkin labels 
$m_i \neq 0$, i.e.~it is specified by a painted Dynkin diagram
\cite{Bordemann:1985xy}. 
\begin{figure}[!t]
\begin{center}
\includegraphics[width=40mm]{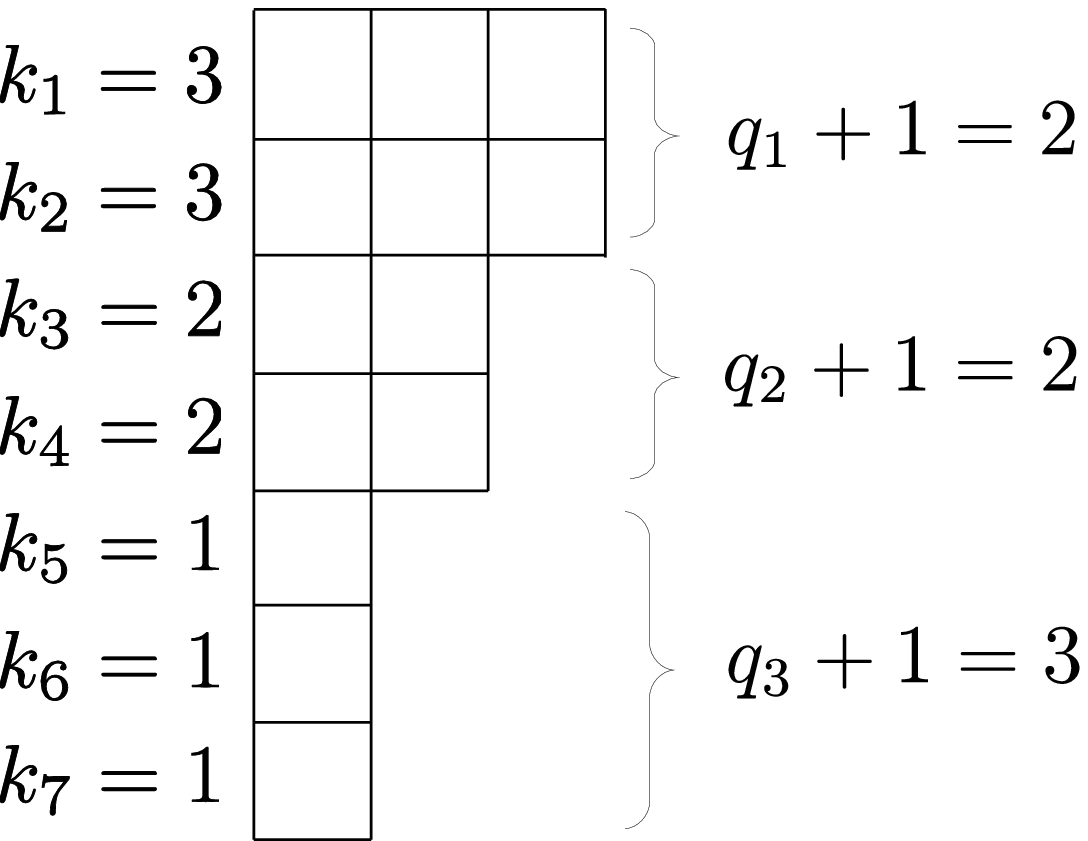} \hs{15}
\includegraphics[width=70mm]{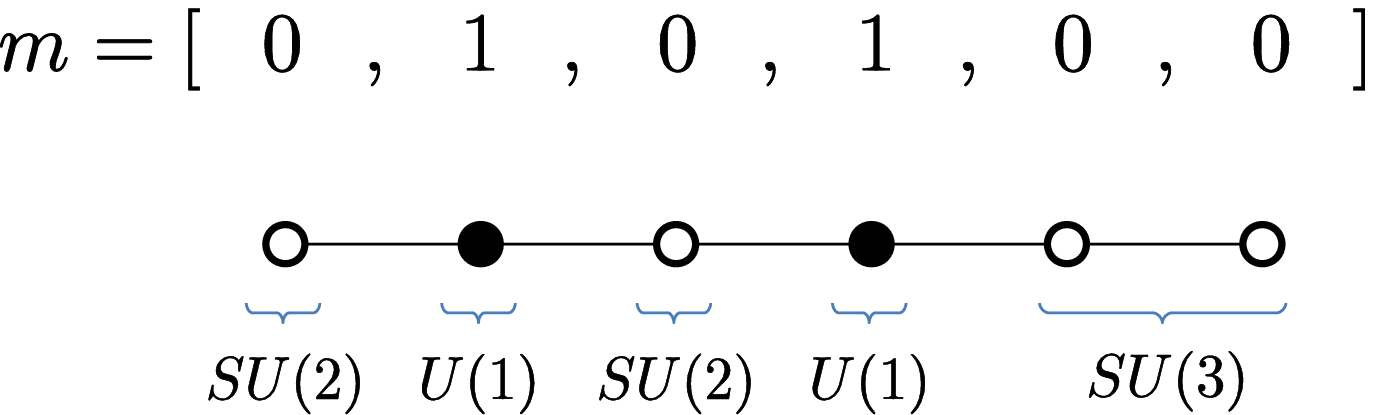}
\caption{{\small An example with $N=7$, 
$m=[0,1,0,1,0,0]$ (and $k=13$), 
$\mathcal M_{\rm orbit} =\frac{SU(7)}{SU(2) \times SU(2) \times SU(3)
    \times U(1)^2}$. 
The black nodes in the Dynkin diagram denote the removed nodes
\cite{Bordemann:1985xy}.}} 
\label{fig:Dynkin}
\end{center}
\end{figure}
Therefore, the irreducible orbits can be written as 
generalized flag manifolds\footnote{These orbits were studied in a non-systematic way in Ref.~\cite{Eto:2004ii}.}
\beq
\mathcal{M}_{\rm orbit} = 
\frac{SU(N)}{SU(q_1 + 1) \times \cdots \times 
SU(q_{p+1} + 1) \times U(1)^p}, 
\label{eq:flag}
\eeq
where $p~(1 \leq p \leq N-1)$ is the number of removed nodes and
$q_i~(i=1,\ldots,p+1)$ is the number of nodes 
in the connected component between the  $(i-1)$-th and $i$-th removed nodes 
(see Fig.\,\ref{fig:Dynkin}). 
The number $p$ is denoted the rank of 
the K\"ahler coset space \eqref{eq:flag}. 
One can also verify that an $H$-transformation on $(\psi,Z)$ 
can indeed be absorbed by $GL(k,\C)$ transformations.

It will now be shown that the irreducible orbits are the fixed-point 
set of the spatial rotation 
\beq
(\psi, Z) ~\rightarrow~ (\psi, e^{i \theta} Z).
\label{eq:rotation}
\eeq
To see this, it is sufficient to check that the highest-weight state
is invariant under the rotation \eqref{eq:rotation}, 
since the $SU(N)$ transformations commute with the spatial rotation. 
One way to show the invariance of the highest-weight state is 
to find a $GL(k,\C)$ transformation which cancels the transformation
\eqref{eq:rotation} on the matrix of Fig.\,\ref{fig:psi_Z}. 
A different, but easier, way is to check the invariance of 
the highest-weight state \eqref{eq:highest} 
under the action of the spatial rotations explicitly. 
Since the generator of the spatial rotation $\hat{J}$ acts on the ground state 
$|0 \rangle \equiv | 0, r_{1}\rangle \otimes \cdots \otimes  | 0, r_{k}\rangle $ and the operators $\hat{a}, \hat{a}^\dagger$ as ($J$ is just a number operator)
\beq
\hat J | 0 \rangle = 0, \qquad 
[\hat J, \hat a_i] = -  \hat a_i, \quad 
[\hat J, \hat a_i^\dagger] =  \hat a_i^\dagger,
\eeq
the highest-weight state \eqref{eq:highest} is an eigenstate of
$\hat{J}$, hence the state transforms as
\beq
| B \rangle ~\rightarrow~ 
\exp\big( i \theta \hat{J} \big) | B \rangle 
= \exp \left(  i \sum_{n=0}^{k_1-1} n \, l_{n+1} \theta \right) 
| B \rangle.
\label{eq:highest_rotation}
\eeq
Since the phase of the state vector is unphysical,
Eq.\,\eqref{eq:highest_rotation} shows that the highest-weight state
is invariant under the spatial rotation. 
Therefore, the irreducible orbits are 
in the fixed-point set of the spatial rotation. 
The inverse also turns out to be true: we can show by using the moduli-matrix formalism that any fixed points of the spatial rotation are
contained in one of the irreducible orbits. 
Therefore, the fixed-point set is precisely 
the disjoint union of the irreducible orbits. 

All this can be seen more explicitly in terms of the original fields. 
The solution $(H,A_\mu)$ to the BPS equation \eqref{eq:bps} 
corresponding to the irreducible orbits can be determined 
from the fact that they are invariant under the spatial rotation
\beq
H(z, \bar z) \rightarrow 
H(e^{-i \theta} z, e^{i \theta} \bar z), \qquad 
A_{\bar z}(z, \bar z) \rightarrow 
e^{i \theta} A_{\bar z}(e^{-i \theta} z, e^{i \theta} \bar z),
\eeq
where $A_{\bar z} = A_1 + i A_2$. 
Let $(H^{(k)},A_\mu^{(k)})$ be the solution of $k$ ANO vortices
situated at the origin $z=0$. 
They transform under the rotation as
\beq
H^{(k)}\left(e^{-i\theta} z, e^{i\theta} \bar{z}\right) 
= e^{-i k \theta} H^{(k)}(z,\bar z), \qquad
A_{\bar z}^{(k)}\left(e^{-i\theta} z, e^{i\theta} \bar{z}\right) 
= A_{\bar z}^{(k)}(z,\bar z), 
\eeq
The solution on the irreducible orbits can be obtained by embedding
the ANO solutions into diagonal components 
\beq
H = U^\dagger \diag\left(H^{(k_1)}, H^{(k_2)}, \cdots, H^{(k_N)}\right) U, \qquad
A_{\bar z} = U^\dag\left(A_{\bar z}^{(k_1)},A_{\bar z}^{(k_2)},\cdots,
A_{\bar z}^{(k_N)}\right) U,
\label{eq:irreducible_sol}
\eeq
where $U \in SU(N)_{C+F}$. 
Note that the sequence of the numbers $\{k_1,k_2,\cdots,k_N\}$ 
can always be reordered as $k_1 \geq k_2 \geq \cdots \geq k_N \geq 0$ 
by using the Weyl group $\mathfrak{S}_N \subset SU(N)_{C+F}$. 
This solution is invariant under the rotation 
since the phase factors of the Higgs fields 
can be absorbed by the following gauge transformation 
\beq
H \rightarrow g H, \quad
A_{\bar z} \rightarrow g A_{\bar z} g^\dagger, \quad
g = U^\dagger \diag\left( e^{-i k_1 \theta}, e^{- i k_2 \theta}, \cdots,
e^{- i k_N \theta}\right) U \in U(N)_{C}. 
\eeq
We can also see that the solution \eqref{eq:irreducible_sol} is
invariant under the same subgroup of $SU(N)$ 
as the state on the irreducible orbit 
specified by the Young tableau with $k_i$ boxes in the $i$-th row.
Therefore, the irreducible orbit with the set of Dynkin labels
$[m_1,\cdots,m_{N-1}]~(m_i = k_i - k_{i+1})$
corresponds to the BPS solutions of the form of
Eq.\,\eqref{eq:irreducible_sol}.

In the next section, we will show that {\it a vortex state at a generic 
point on the moduli space is given by a linear superposition of
vectors corresponding to various irreducible representations. }
Furthermore, in Section\,\ref{sec:kahler}, 
metrics for all irreducible $SU(N)$ orbits will be obtained 
by assuming that the metrics are K\"ahler and 
isometric under the $SU(N)$ action.

\section{$SU(N)$ decomposition of general $k$ vortex states}
\label{sec:examples}

In this section we solve the constraints \eqref{eq:constraint2} and \eqref{eq:constraint3} in order to find the $SU(N)$ property of a general $k$-winding 
vortex. 
The cases of $k=1$, $2$ and $3$ are solved concretely; a general recipe for the solution will be given, valid for any $N$ and for any winding number $k$.
A particular attention will be paid to the vortices with coincident centers. The results of these analyses provide the $SU(N)$ decomposition rule for a generic vortex state of a given winding number.

\subsection{ $k=1$ vortices \label{sec:k=1}}

$k=1$ is a trivial example. In this case, we have 
\beq
\sigma_1 = z_1, \qquad
\left| B \right> = \sum_{r=1}^N \phi_r \left| z_1, r \right>.
  \label{eq:k=1-B}
\eeq
There is no nontrivial constraint, so that the moduli space is
\beq
{\cal M}^{k=1} 
~=~ \mathbb{C} \times \mathbb{C}P^{N-1}
~\simeq~ \mathbb{C} \times \frac{SU(N)}{SU(N-1) \times U(1)}.
\label{eq:k=1}
\eeq
As $\left| B \right>$ is in the fundamental representation of $SU(N)$, 
the orientational moduli space is given by the orbit of 
a vector in the fundamental representation.

\subsection{Solution of the constraints for $k=2$ \label{sec:k=2}}

This is the first case with nontrivial constraints.

\subsubsection*{$k=2$ $U(N)$ vortices}

With coordinates $\sigma_1=z_1+z_2\in \mathbb C$ and 
$\sigma_2=z_1z_2 \in \mathbb C$, the linear constraints
\eqref{eq:constraint2} in this case are given by 
\beq
(\hat a_1+\hat a_2) \left| B\right> = \sigma_1\left| B\right>,\quad 
\hat a_1 \hat a_2 \left| B\right> = \sigma_2 \left| B\right>, 
\label{eq:sol-B}
\eeq
which are equivalent to the following equations 
for the baryonic invariants
\beq
B^{n+1}_{\hs{2}r}{}^{m}_{s}+B^{n}_{r}{}^{m+1}_{\hs{2}s}
=\sigma_1B^{n}_{r}{}^{m}_{s},\quad 
B^{n+1}_{\hs{2}r}{}^{m+1}_{\hs{2}s}=\sigma_2B^{n}_{r}{}^{m}_{s}.
\eeq 
In Section~\ref{sec:sepa}, we have seen that 
the solution can be expressed by the coherent states for separated vortices. 
Let us see what happens to the coherent states in the coincident limit.  
In the case of $k=2$, 
the coherent state representation of the solution is given by 
\beq
\left| B \right> &=& \frac{1}{2} \tilde{B}_{r_1 r_2} 
\Big( \left| z_1 , r_1 \right> \otimes 
\left| z_2 , r_2 \right> - \left| z_2 , r_2 \right> \otimes 
\left| z_1 , r_1 \right> \Big). 
\label{eq:k=2_coherent}
\eeq
It is convenient to decompose $\tilde B_{r_1 r_2}$ 
into the irreducible representations of $SU(N)$
\beq
\tilde A_{r_1 r_2} \equiv \frac{\tilde B_{r_1 r_2} - \tilde B_{r_2 r_1}}{2}, \qquad
\tilde S_{r_1 r_2} \equiv \frac{\tilde B_{r_1 r_2} + \tilde B_{r_2 r_1}}{2}.
\eeq
Then, the solution can be rewritten as
\beq
\left| B \right> = \left[\tilde{A}_{r_1 r_2} 
\cosh \frac{(z_1-z_2)(\hat{a}_1^\dagger-\hat{a}_2^\dagger)}{2} 
+\tilde{S}_{r_1 r_2} 
\sinh \frac{(z_1-z_2)(\hat{a}_1^\dagger-\hat{a}_2^\dagger)}{2} \right] 
\left| \frac{\sigma_1}{2}, r_1 \right> \otimes 
\left| \frac{\sigma_1}{2}, r_2 \right>, \nonumber
\eeq
where $\sigma_1 = z_1 + z_2$. 
If we na\"{i}vely take the coincident limit $z_2 \rightarrow z_1$, 
the symmetric part drops out 
\beq
\left| B \right> &\rightarrow& 
\tilde{A}_{r_1 r_2} \left| \frac{\sigma_1}{2}, r_1 \right> \otimes 
\left| \frac{\sigma_1}{2}, r_2 \right>. 
\eeq
Although this state satisfies the constraint \eqref{eq:sol-B}, 
this is not the most general solution in the coincident case. 
To obtain the correct expression for the most general solution, 
let us redefine 
\beq
A_{r_1 r_2} \equiv \tilde{A}_{r_1 r_2}, \qquad 
S_{r_1 r_2} \equiv \frac{z_1-z_2}{2} \, \tilde{S}_{r_1 r_2}.
\eeq
Then, the solution \eqref{eq:k=2_coherent} can be rewritten as 
\beq
\left| B \right> = 
\sum_{n=0}^\infty \frac{1}{(2n)!} w^n (\hat{a}_1^\dagger-\hat{a}_2^\dagger)^{2n} 
\left[ A_{r_1 r_2} + \frac{1}{2n+1} S_{r_1 r_2}(\hat{a}_1^\dagger-\hat{a}_2^\dagger) \right] 
\left| \frac{\sigma_1}{2}, r_1 \right> \otimes 
\left| \frac{\sigma_1}{2}, r_2 \right>,
\label{eq:k=2_general_sol}
\eeq
where we have introduced a square of the relative position as
\beq
w ~\equiv~ \frac{\sigma_1^2}{4} - \sigma_2 
~=~ \frac{(z_1-z_2)^2}{4}. 
\eeq
In this expression, it is obvious that the symmetric part 
also survives in the coincident limit $w \rightarrow 0$
\beq
\left| B \right> &\rightarrow& 
\left[ A_{r_1 r_2} + S_{r_1 r_2} (\hat{a}_1^\dagger-\hat{a}_2^\dagger) \right] 
\left| \frac{\sigma_1}{2}, r_1 \right> \otimes 
\left| \frac{\sigma_1}{2}, r_2 \right>.
\eeq
Therefore, Eq.\,\eqref{eq:k=2_general_sol} is 
the most general form of the solution 
which is valid also in the coincident limit.
The symmetric and anti-symmetric tensors $S_{rs}$ and $A_{rs}$ 
in Eq.\,\eqref{eq:k=2_general_sol} are 
the well-defined coordinates of the vector space $W(\sigma_i)$ 
for arbitrary values of $\sigma_i$.   
Clearly these correspond to the decomposition 
of the tensor product $\mathbf{N} \otimes \mathbf{N}$ 
into irreducible representations of $SU(N)$. 
{\it A generic point on the moduli space is described by a
superposition of the states belonging to different irreducible representations. }

In terms of $A_{rs}$ and $S_{rs}$, 
the baryonic invariants can be read off from the solution using
\eqref{eq:state_to_B}
\beq
B^{00}_{rs}=A_{rs}, \quad 
B^{10}_{rs}=S_{rs}+\frac{\sigma_1}2 A_{rs}, \quad 
B^{11}_{rs}=\sigma_2 A_{rs},\quad \cdots ,
\eeq
and hence, the Pl\"ucker conditions \eqref{eq:constraint3}, 
which are the remaining constraints, can be rewritten as 
\beq
A_{pq}A_{rs} + A_{pr}A_{sq} + A_{ps}A_{qr} &=& 0,
\label{eq:pl1}\\
A_{pq}S_{rs} + A_{rp}S_{qs} + S_{ps}A_{qr} &=& 0,
\label{eq:pl2}\\
 w\,  A_{pq}A_{rs} + S_{pr}S_{qs} - S_{ps}S_{qr} &=& 0.
\label{eq:pl3}
\eeq
By these constraints, the moduli space of two vortices 
is embedded into $\C^2 \times \C P^{N^2-1}$ 
which is parametrized by independent coordinates 
$\{\sigma_1,\sigma_2,A_{rs},S_{rs}\}$. 

Now, let us look into two different subspaces 
corresponding to the irreducible $SU(N)$ orbits.  
They are obtained by setting
1) $S_{rs}=0$,~$A_{rs} \neq 0$ and 2) $A_{rs} = 0$,~$S_{rs}\neq0$. 

1) Consider first the subspace with $S_{rs}=0$.
Eq.\,\eqref{eq:pl3} allows $S_{rs}=0$ only 
in the coincident case $w=0$. 
Note that Eq.\,\eqref{eq:pl2} is automatically satisfied by
$S_{rs}=0$, and that Eq.\,\eqref{eq:pl1} gives the ordinary Pl\"ucker
conditions which embed the complex Grassmannian $Gr_{N,2}$ into a
complex projective space 
$\mathbb{C}P^{N(N-1)/2-1} \simeq \{A_{pq}\}/\!\!/\mathbb{C}^*$. 
We find therefore that the subspace $S_{rs}=0$ is:
\beq
{\cal M}^{{\def\YGbox{4} {\Young[0]{11}}}} ~\cong~ 
\mathbb{C} \times  Gr_{2,N} ~\cong~ 
\mathbb{C} \times \frac{SU(N)}{SU(2) \times SU(N-2) \times U(1)}.
\eeq 
According to the results in the previous section, this is the
irreducible $SU(N)$ orbit for ${\def\YGbox{4} {\Young[0]{11}}}$~.   

2) In the other subspace characterized by $A_{rs}=0$, 
we have a nontrivial constraint $S_{pr}S_{qs} = S_{ps} S_{qr}$. 
The general solution is
\beq
S_{rs} = \phi_r \phi_s, \qquad 
\phi_r \in \C^{N}.
\eeq
Here $\phi_r$ is nothing but the orientation vector given 
in Eq.\,\eqref{eq:k=1-B}, 
so $S_{rs} = \phi_r\phi_s$ corresponds to 
the $k=2$ vortices with parallel orientations.
The corresponding moduli subspace is given by
\beq
{\cal M}^{{\def\YGbox{4} {\Young[0]{2}}}} ~\cong~ 
\mathbb{C}^{2} \times \mathbb{C}P^{N-1} ~\cong~ 
\mathbb{C}^{2} \times \frac{SU(N)}{SU(N-1) \times U(1)},
\label{eq:k=2p}
\eeq
which is indeed the other irreducible orbit, extended for generic $w$.  
We have thus identified the two moduli subspaces, 
the irreducible $SU(N)$ orbits of anti-symmetric and symmetric
representations, respectively. 
They correspond to the vortex states in Eq.\,\eqref{eq:k=2_general_sol} 
without the second or the first term, respectively. 
The generic vortex state \eqref{eq:k=2_general_sol} is 
a linear superposition of these two states.

Note that in some cases the orbits of different representations are 
described by the same coset manifold. 
For example, both $\square$\ and $\square\!\square$  
are given by ${\mathbb C}P^{N-1}$, see Eqs.\,\eqref{eq:k=1} and
\eqref{eq:k=2p}. 
As we shall see in Section\,\ref{sec:kahler}, however, 
the K\"ahler class completely 
specifies the representations and distinguishes the orbits 
belonging to different representations \footnote{Except for the cases of pairs of conjugate representations.
They are found to be described by the same K\"ahler metric, i.e., by the same low-energy effective action.  See Subsection~\ref{sec:conj} below. }.  

\subsubsection*{More on $k=2$ coincident $U(2)$ vortices}

Let us study $k=2$ vortices in the $U(2)$ case in some more detail 
by looking at another slice of the moduli space.
This case in particular has been studied in the
Refs.~\cite{Hashimoto:2005hi,Auzzi:2005gr,Eto:2006cx,Eto:2006db,Eto:2006dx,Auzzi:2010jt}.
In this case, there exist only a singlet $A_{12}$ 
and a triplet $\{S_{11},S_{12},S_{22}\}$ of $SU(2)$.

\begin{figure}[!t]
\begin{center}
\includegraphics[height=7cm]{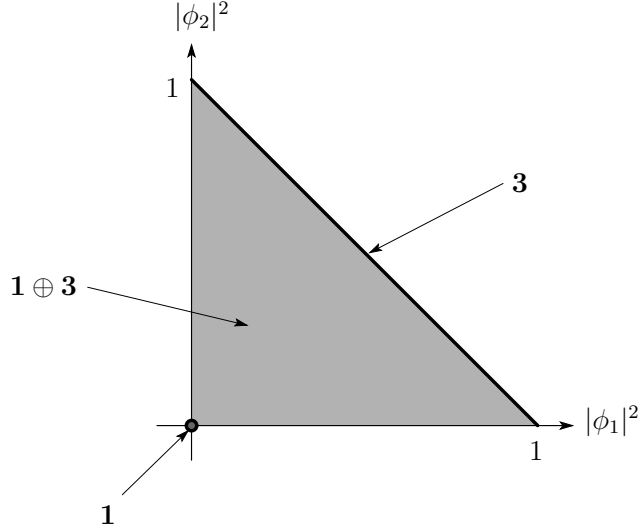}
\caption{{\small $W\mathbb{C}P^2_{(2,1,1)}$ in the gauge 
$2|\eta|^2 + |\phi_1|^2 + |\phi_2|^2 = 1$. 
The diagonal edge corresponds to the triplet state ${\bf 3}$ 
and the origin to the singlet state ${\bf 1}$. The bulk is a
nontrivial superposition of ${\bf 1}$ and ${\bf 3}$. 
The diagonal edge and the origin are the only irreducible orbits in this system.}} 
\label{fig:wcp2}
\end{center}
\end{figure}

Among the constraints \eqref{eq:pl1}--\eqref{eq:pl3}, 
the only nontrivial one is 
\beq
w\,  (A_{12})^2 + S_{11}S_{22} - (S_{12})^2 = 0.
\eeq
Let us consider the moduli space of coincident vortices 
which corresponds to the subspace $w = 0$. 
In this case, the above constraint is solved by 
$S_{rs} = \phi_r \phi_s$ again. 
Now, the moduli subspace is parametrized by 
the center of mass position $z_0=\frac{\sigma_1}{2}$ and 
$\{\eta,\phi_1, \phi_2\}$ with $\eta \equiv A_{12}$. 
Thus, the vortex state is given, without constraints, by 
\beq
\left| B\right>_{w=0} &=& 
\eta\left|z_0\right>_{\bf 1}
+\sum_{r,s=1}^2\phi_r\phi_s \left|z_0;r,s\right>_{\bf 3},
\eeq
where the singlet $\left|z_0\right>_{\bf 1}$ and the triplet 
$\left|z_0;r,s\right>_{\bf 3}$ are given by
\beq
\left|z_0\right>_{\bf 1} &\equiv& 
\left| z_0,1\right>\otimes 
\left| z_0,2\right>
-\left| z_0,2\right>\otimes 
\left| z_0,1\right>,\\
\left|z_0;r,s\right>_{\bf 3} &\equiv& 
(\hat{a}_1^\dagger-\hat{a}_2^\dagger)
\Big( \left| z_0,r\right>\otimes \left| z_0,s\right>+\left| z_0,s\right>\otimes \left| z_0,r\right> \Big). 
\eeq
Note that the $\mathbb{C}^*\subset GL(k,\mathbb C)$ acts  as
\beq
\{\eta,\phi_1,\phi_2\} \sim 
\{\lambda^2 \eta,\lambda \phi_1,\lambda\phi_2\},\quad 
\lambda \in \mathbb{C}^*. \label{eq:phase-trf}
\eeq
Hence the moduli subspace for the two coincident vortices is 
found to be the two dimensional weighted projective space 
with the weights $(2,1,1)$ 
\beq
{\cal M}_{k=2}^{\text{coincident}} ~\cong~ 
\mathbb{C} \times W\mathbb{C}P^2_{(2,1,1)} ~\cong~ 
\mathbb{C} \times \frac{\mathbb{C}P^2}{\mathbb Z_2}.
\eeq 
This is exactly the result obtained previously
\cite{Auzzi:2005gr,Eto:2006cx}. 
Although this might be seen as just a reproduction of an old result, 
there is a somewhat new perspective on 
the irreducible representation of $SU(2)$. 
Here we would like to stress again that $A_{12}=\eta$ is
the singlet while $S_{rs}=\phi_r\phi_s$ is the triplet.  
{\it Together} they form the coordinate of
$W\mathbb{C}P^2_{(2,1,1)}$. 
In Fig.\,\ref{fig:wcp2}, we show the space $W\mathbb{C}P^2_{(2,1,1)}$ 
in the $|\phi_1|^2$--$|\phi_2|^2$ plane with a natural metric given  
by $2|\eta|^2 + |\phi_1|^2 + |\phi_2|^2 = 1$. 
The states ${\bf 3}$ and ${\bf 1}$ live on the boundaries of 
$W{\mathbb C}P^2_{(2,1,1)}$; the points in the bulk of
$W\mathbb{C}P^2_{(2,1,1)}$ are described by the superposition 
${\bf 1}\oplus{\bf 3}$.   

In Appendix \ref{sec:WCP2} 
we discuss possible metrics on $W\mathbb{C}P^2_{(2,1,1)}$ 
and show that independently of the choice of the metric, 
they indeed yield at the diagonal edge of 
Fig.\,\ref{fig:wcp2} the Fubini-Study metric with the same K\"ahler
class on ${\mathbb C}P^1$. 

\subsection{Solution for the $k=3$ coincident vortices \label{sec:k=3}}

In this section, we consider $k=3$ vortices sitting 
all at the origin, $\sigma_1 = \sigma_2 = \sigma_3 = 0$ ($z_1=z_2=z_3=0$).   (The $k=3$ vortex solutions of more general types
-- with generic center positions -- will be discussed in Appendix~\ref{appendix:k=3}.)
The constraint \eqref{eq:constraint2} reduces to 
\beq
&\hat{a}_1 \hat{a}_2 \hat{a}_3 \left| B \right> = 0, \qquad
(\hat{a}_1 \hat{a}_2 + \hat{a}_2 \hat{a}_3 + \hat{a}_3 \hat{a}_1) 
\left| B \right> = 0, \qquad
(\hat{a}_1 + \hat{a}_2 + \hat{a}_3) \left| B \right> = 0,
\eeq
which lead to $(\hat{a}_i)^3 \left| B \right> = 0$ for $i = 1, 2, 3$.  
Taking into account the anti-symmetry condition \eqref{eq:anti-symmetric}, 
we obtain the following solution to the constraints 
(see Appendix \ref{appendix:k=3})
\beq
\left| B \right> &=& \Big[ 
A_{r_1r_2r_3} + \left( 
X^1_{r_1 r_2 r_3} \hat{a}_1^\dagger + 
X^2_{r_1 r_2 r_3} \hat{a}_2^\dagger + 
X^3_{r_1 r_2 r_3} \hat{a}_3^\dagger \right) \notag \\
&{}& \quad
- \frac{1}{2} \left( 
Y^1_{r_1r_2r_3} (\hat{a}_2^\dagger - \hat{a}_3^\dagger)^2 + 
Y^2_{r_1r_2r_3} (\hat{a}_1^\dagger - \hat{a}_3^\dagger)^2 +
Y^3_{r_1r_2r_3} (\hat{a}_1^\dagger - \hat{a}_2^\dagger)^2 \right)
\notag \\ 
&{}& \quad - 
\frac{1}{2} S_{r_1r_2r_3} 
( \hat{a}_1^\dagger - \hat{a}_2^\dagger ) 
( \hat{a}_2^\dagger - \hat{a}_3^\dagger ) 
( \hat{a}_3^\dagger - \hat{a}_1^\dagger ) \Big] 
\left|0,r_1 \right> \otimes 
\left|0,r_2 \right> \otimes 
\left|0,r_3 \right>, \label{eq:vortexstatek3}
\eeq
where $Y^i_{r_1r_2r_3}~(i=1,2,3)$ and $X^i_{r_1r_2r_3}~(i=1,2,3)$ 
are tensors satisfying
\beq
Y^1_{r_1r_2r_3}+Y^2_{r_1r_2r_3}+Y^3_{r_1r_2r_3}=0, \qquad
X^1_{r_1r_2r_3}+X^2_{r_1r_2r_3}+X^3_{r_1r_2r_3}=0.
\eeq
The tensors $S,Y,X,A$ have the following index structures
\beq
S_{r_1 r_2 r_3} &=& S_{r_{\rho(1)} r_{\rho(2)} r_{\rho(3)}}, \\ 
Y^i_{r_1 r_2 r_3} &=& {\rm sign} (\rho) Y^{\rho(i)}_{r_{\rho(1)} r_{\rho(2)} r_{\rho(3)}}, \\
X^i_{r_1 r_2 r_3} &=& {\rm sign} (\rho) X^{\rho(i)}_{r_{\rho(1)} r_{\rho(2)} r_{\rho(3)}}, \\
A_{r_1 r_2 r_3} &=& {\rm sign} (\rho) A_{r_{\rho(1)} r_{\rho(2)} r_{\rho(3)}},
\eeq
where $\rho$ denotes elements of the symmetric group $\mathfrak{S}_3$. 
The first and last equation show that 
$S_{r_1r_2r_3}$ and $A_{r_1r_2r_3}$ are totally symmetric and
anti-symmetric, respectively. 
The second (third) equation indicates that only one of 
$Y^1,Y^2,Y^3~(X^1,X^2,X^3)$ is independent.
Hence we arrive at a natural correspondence between the baryons and
the Young tableaux as 
\beq
\def\YGbox{10}
A_{r_1r_2r_3}: \Young[-1]{111}\,,\quad
X^i_{r_1r_2r_3}: \Young[-0.5]{21}\,,\quad
Y^i_{r_1r_2r_3}: \Young[-0.5]{21}\,,\quad
S_{r_1r_2r_3}: \Young[0]{3}\,.
\label{eq:B's}
\eeq
This looks perfectly consistent 
with the standard decomposition of 
$\square \otimes \square \otimes \square$. 

\begin{figure}[!t]
\begin{center}
\begin{tabular}{ccc}
\includegraphics[width=47mm]{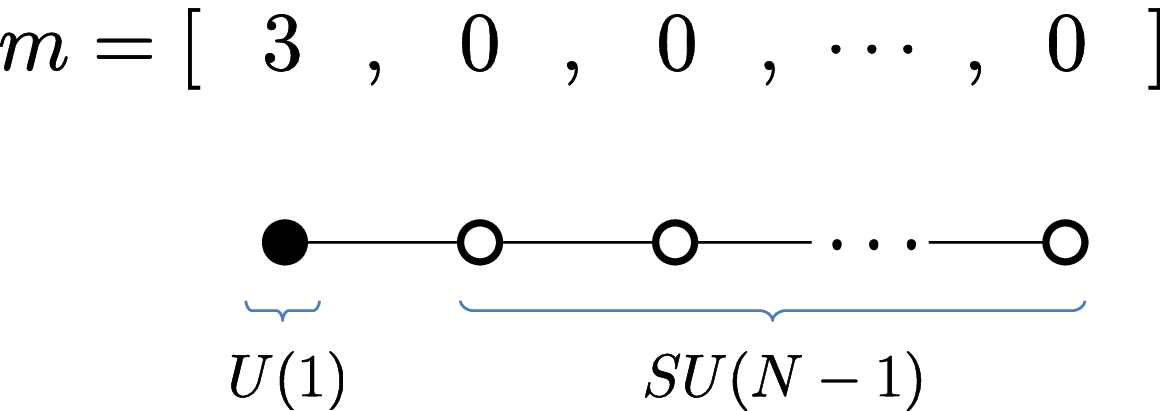} \hs{1}&\hs{1}
\includegraphics[width=47mm]{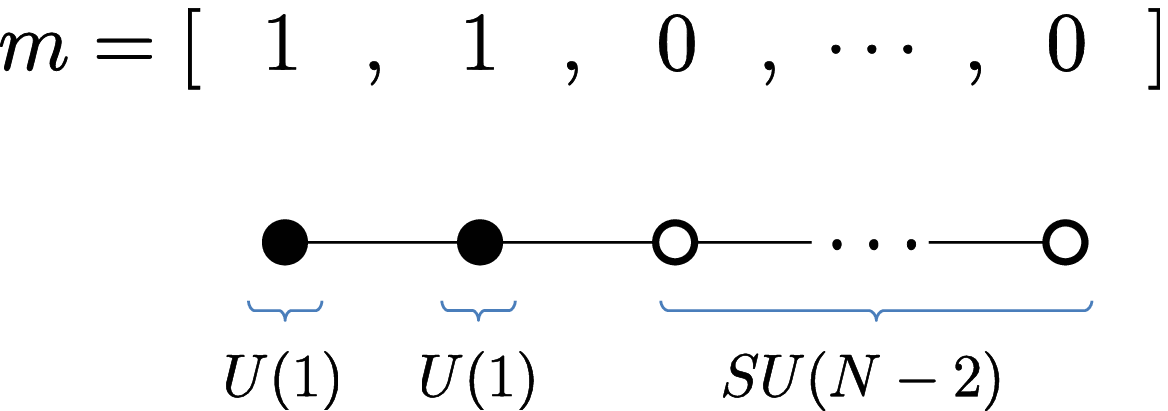} \hs{1}&\hs{1} 
\includegraphics[width=57mm]{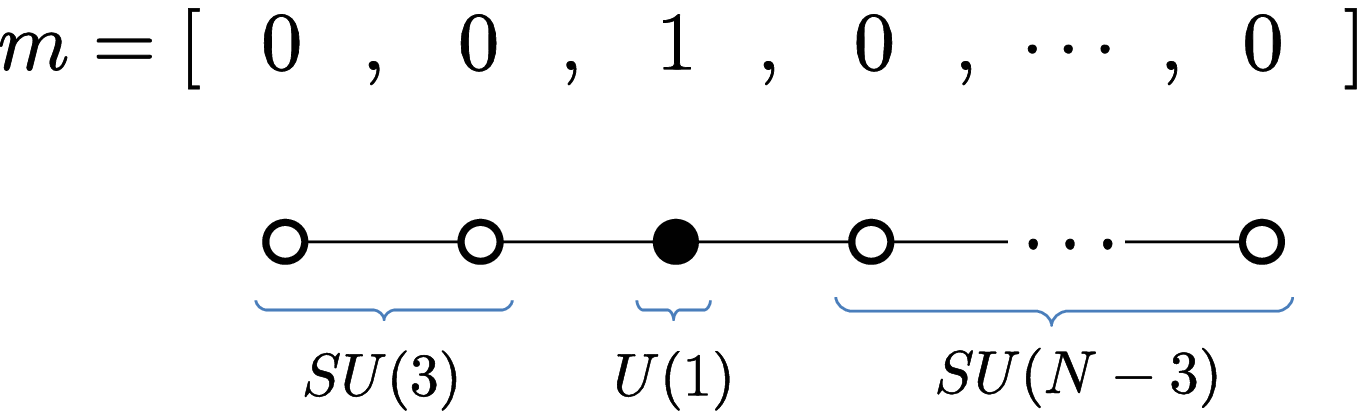} \\
$S$ : \Young[0]{3} & 
$X$ : \Young[-0.5]{21} & 
$A$ : \Young[-1]{111}
\end{tabular} 
\caption{The irreducible orbits in the moduli space of $k=3$ vortices.}
\label{asin}
\end{center}
\end{figure}

Actually this is not quite straightforward, 
and this example nicely illustrates the subtlety alluded in the Introduction.   
As we have seen in the previous section, 
there is a one-to-one correspondence between 
the highest-weight states of the baryons $\left| B \right>$ and 
the Young tableaux with $k$ boxes of a definite type. 
This means that there is only one vortex state of highest weight, 
corresponding to the mixed-symmetry Young tableau\footnote{In
  contrast to the standard composition-decomposition rule for three
  distinguishable objects in the ${\bf N}$ representation, {\it  two}
  inequivalent highest weight states in the same irreducible
  representation, described by the same mixed-type Young tableau, will
  appear. This is not so for our $k$ vortices. }. However,  we seem to
have $Y$ and $X$ in \eqref{eq:B's}, both of which correspond to the
same Young 
tableau. This apparent puzzle is solved by looking at the following 
Pl\"ucker relation rewritten in terms of $S,X,Y,A$ 
\beq
(Y^1_{rst})^2 = - S_{rsr} X^1_{tst} - X^1_{srs} S_{rtt} + X^1_{srt}
S_{rts}, \quad
\mbox{(no sum over $r,s,t$)}\;,
\label{eq:k=3_Plucker}
\eeq 
which shows that the tensor $Y$ is determined in terms of the others up to a sign. 
This implies that no solution to Eq.\,\eqref{eq:k=3_Plucker} of ``pure $Y$'' type, i.e., 
with $Y \neq 0,  \, A = S= X= 0$, exists. 
Hence we have verified the one-to-one correspondence between the
highest-weight baryon states $\left| B \right>$ and the Young
tableaux, as in Figure~\ref{asin}. 

By setting two among $S$, $X$ or $A$ to be zero, we obtain the
corresponding $SU(N)$ irreducible orbits, which can be immediately
read off from the Young tableaux as (for $N \ge k=3$) 
\beq
{\cal M}^{S} &\cong& \frac{SU(N)}{SU(N-1) \times U(1)} ~\cong~ 
\mathbb{C}P^{N-1}, \\
{\cal M}^{X} &\cong& \frac{SU(N)}{SU(N-2) \times U(1)^2},\\
{\cal M}^{A} &\cong& \frac{SU(N)}{SU(3) \times SU(N-3) \times U(1)}
~\cong~ 
Gr_{N,3}.
\eeq

Due to the existence of $Y$, the whole subspace with $\sigma_i=0$ is
more complicated than the $k=2$ case. 
The simplest nontrivial case $N=2$ ($SU(2)$ global symmetry)  somewhat enlightens our understanding. 
In that case, $A$ is identically zero and the following
parametrization using the coordinates  
$\{\eta, \xi^1, \xi^2, \phi_1, \phi_2 \} \in \mathbb C^5$
\beq
X^1_{r12} = \epsilon_{rs} \xi^s, \quad 
Y^1_{r12} = \eta \, \phi_r, \quad 
S_{rst} = \phi_r \phi_s \phi_t, \quad r,s,t=1,2
\eeq
solves all of the Pl\"ucker relations except for 
\beq
\eta^2 = \xi^r \phi_r.
\eeq
Therefore, $\eta$ is a locally dependent coordinate. 
Since the equivalence relation is 
\beq
\{\xi^r, \eta, \phi_r \} ~\simeq~ 
\{\lambda^3 \xi^r , \lambda^2 \eta, \lambda \phi_r \},
\eeq
the moduli space in this case is a hypersurface in 
$W\mathbb CP^4_{(3,3,2,1,1)}\simeq \mathbb CP^4/\mathbb Z_3$.
The irreducible orbits corresponding to $S$ and $X$ are the subspaces 
obtained by setting $\xi^r=0$ or $\phi_r=0$, respectively. 
Both of them are isomorphic to 
\beq
{\cal M}^{S} &\cong& 
{\cal M}^{X} ~\cong~ \frac{SU(2)}{U(1)} ~\cong~ 
\mathbb{C}P^{1}.
\eeq
According to the results of the next section, however, they are
characterized by the different K\"ahler classes while their K\"ahler
potentials are given by 
\beq
K\simeq \left\{
\begin{array}{cc}
  3 r \log|\phi_r|^2   & {\rm as~} |\xi_i|^2\to 0\\
r \log|\xi^r|^2  & {\rm as~} |\phi_i|^2\to 0\\
\end{array}\right. .
\eeq

\subsection{Generalization to arbitrary winding number}

In this section, we comment on a generalization to 
the case of an arbitrary winding number $k$. 
As we have seen in the $k=2,3$ cases, 
the coherent states \eqref{eq:coherent} become insufficient 
to describe the general solution to the constraint \eqref{eq:constraint2} 
when two or more vortex centers coincide. 
The procedure to obtain the general solution for $k=3$ vortices 
can be generalized to the case of arbitrary $k$ as follows. 
Let $\left| S ; r_1,\cdots,r_k ; \{z_i\} \right>$ be 
the following linear combination of the coherent states
\beq
\left| S ; r_1,\cdots,r_k ; \{z_i\} \right> \equiv \frac{1}{k! \, \Delta}
\sum_{\rho \in \mathfrak S_k} {\rm sign}(\hat \rho) \ \hat \rho \,
\hat v \, \hat \rho^{-1} \left|0, r_1 \right> \otimes \cdots \otimes \left| 0 ,r_k \right>, \label{welldef}
\eeq
where the polynomial $\Delta$ and 
the operators $\hat v$ are defined by
\beq
\Delta \equiv \prod_{I>J} ( z_I - z_J ), \hs{10} 
\hat v \equiv \exp\left(\sum_{i=1}^k z_i a_i^\dag\right) \;;  
\eeq
 $\hat \rho \, \hat v \, \hat \rho^{-1}$ then reads
\beq
\hat \rho \, \hat v \, \hat \rho^{-1} = \exp \left( z_1 \hat
a_{\rho^{-1}(1)}^\dagger + z_2 \hat a_{\rho^{-1}(2)}^\dagger + \cdots + z_k \hat
a_{\rho^{-1}(k)}^\dagger \right). 
\eeq
This state vector (\ref{welldef}) is a solution of the constraint \eqref{eq:constraint2} 
which is well-defined even in the coincident limit $z_I \rightarrow z_J$:  
\beq
\left| S ; r_1,\cdots,r_k ; \{z_i\}\right> \rightarrow \Delta(\hat a_1^\dagger, \cdots, \hat a_k^\dagger) \left|0,r_1 \right> \otimes \cdots \otimes \left|0, r_k \right>.
\eeq
Other well-defined solutions can be obtained 
by acting with polynomials of annihilation operators $\hat a_i$ 
on $\left| S ; r_1,\cdots,r_k ; \{z_i\}\right>$. 
The linearly independent solutions are generated by 
the polynomials $h_i(\hat a_1, \cdots, \hat a_k)$ satisfying 
the following property\footnote{The conditions  (\ref{equival})  can be written in an alternative, equivalent form 
$P(\partial_1, \cdots, \partial_k) h_i(\eta_1,\cdots,\eta_k) =0$,   where
  $\partial_i\equiv {\partial}/{\partial \eta_i}$.} for arbitrary symmetric polynomials
$P$: 
\beq
\left< 0 \right| 
h_i(\hat a_1, \cdots, \hat a_k) \,
P(\hat a_1^\dagger, \cdots, \hat a_k^\dagger) = 0,  \label{equival}
\eeq
where $\left<0\right|\equiv \left<0,r_1\right|\otimes\cdots\otimes
\left<0,r_k\right|$. 
Such polynomials $h_i(\hat a_1, \cdots, \hat a_k)$ span 
a $k!$-dimensional vector space $H$ 
on which the symmetric group $\mathfrak S_k$ acts
linearly\footnote{The representation of $H$ is isomorphic to the
  regular representation of $\mathfrak S_k$.}
\beq
\hat \rho \, h_i(\hat a_1, \cdots \hat a_k) \hat \rho^{-1} ~=~ h_i(\hat a_{\rho^{-1}(1)}, \cdots \hat a_{\rho^{-1}(k)}) ~=~ g_i{}^j(\rho) \, h_j(\hat a_1, \cdots \hat a_k),
\eeq
where $g_i{}^j(\rho)$ is a matrix corresponding to 
the transformation $\rho \in \mathfrak S_k$. 
By using a linearly independent basis $\{h_i \}$, 
the general solution to Eq.\,\eqref{eq:constraint2} 
can be written as a superposition of $h_i \left|S;r_1\cdots r_k \right>$
\beq
\left| B \right> = \sum_{r_1, \cdots r_k} \sum_{i=1}^{k!} X^i_{r_1
  \cdots r_k} h_i (\hat a_1 , \cdots , \hat a_k ) \left| S; r_1,
\cdots, r_k ; \{z_i\} \right>.
\label{eq:general}
\eeq 
Since $\left| S; r_1, \cdots, r_k ; \{z_i\} \right>$ is well-defined 
for arbitrary vortex positions,
this expression of the general solution is valid 
even in the coincident limit. 
Taking into account the constraint Eq.\,\eqref{eq:constraint1}, 
we find that $X^i_{r_1 \cdots r_k}$ should have the following index structure
\beq
X^i_{r_1 \cdots r_k} = X^i_{r_{\rho^{-1}(1)} \cdots r_{\rho^{-1}(k)}} 
g_i{}^j(\rho), \hs{10} \mbox{for all $\rho \in \mathfrak S_k$}.
\label{eq:constraint_X}
\eeq
This condition reduces the number of degrees of freedom 
to $N^k = {\rm dim} \, W(\sigma_i)$. 
Since Eq.\,\eqref{eq:coherent} and Eq.\,\eqref{eq:general} 
are related by the change of basis 
from coherent states to $h_i \left|S;r_1,\cdots,r_k ; \{z_i\}\right>$, 
the coordinates $X^i_{r_1\cdots r_k}$ can be obtained 
from $\tilde B_{r_1 \cdots r_k}$ by a linear coordinate transformation 
with $z_I$-dependent coefficients. 
Therefore, it is obvious that $X^i_{r_1\cdots r_k}$ transforms 
under $SU(N)$ as a multiplet in the direct product representation 
$\otimes_{i=1}^k \mathbf N$. 
We can also confirm this fact by decomposing 
the $k!$-dimensional vector space $H$ into 
the  irreducible representations of the symmetric group $\mathfrak S_k$. 
They are classified by the standard Young tableaux with $k$ boxes 
(Young tableaux with increasing numbers in each row and column)  
and correspondingly, the set of the coefficients $\{ X^i_{r_1 \cdots r_k} \}$ 
can also be decomposed into subsets classified by the standard Young tableaux. 
 Eq.\,\eqref{eq:constraint_X} then tells us that 
the subset of $X^i_{r_1 \cdots r_k}$ 
for each irreducible representation of $\mathfrak S_k$
forms a multiplet in the irreducible representation of $SU(N)$ specified by 
the corresponding Young tableau.

Finally, the remaining constraint \eqref{eq:constraint3} 
can be rewritten by using the relation \eqref{eq:state_to_B} 
to quadratic constraints for $X^i_{r_1 \cdots r_k}$, 
which give the vortex moduli space as a subspace in $\C^k \times \C P^{N^k}$. 


\section{K\"ahler potential on irreducible $SU(N)$ orbits}
\label{sec:kahler}

In this section we will obtain the metric on each of the irreducible
orbits inside the vortex moduli space ${\cal M}_k$ by use of a
symmetry argument.  We only use the fact that the metric of the whole
vortex moduli space is K\"ahler and has an $SU(N)$ isometry. 

One of the most important characteristics of non-Abelian vortices is 
that they possess internal orientational moduli. 
These arise when the vortex configuration breaks 
the $SU(N)_{\rm C+F}$ symmetry to its subgroup $H \subset SU(N)$.  
For a single vortex, it is broken to $SU(N-1)\times U(1)$ 
and the moduli space is homogeneous. 
On the other hand, the moduli space for multiple vortices, i.e.~$k>1$,
is not homogeneous and has some anisotropic directions 
(even if we restrict ourselves to consider 
the subspace of coincident vortices). 
Consequently, the shape of the metric at generic points cannot be
determined from the symmetry alone. 
The metric is not isometric along such a direction, 
and the isotropic subgroup $H$ (and the orbit $SU(N)/H$) 
can change as we move along
such a direction in ${\cal M}_k$.\footnote{This usually occurs in
  supersymmetric theories with spontaneously broken global symmetries
  and is called the supersymmetric vacuum alignment
  \cite{Kotcheff:1988ji}. This phenomenon was discussed for
  non-Abelian vortices in Ref.~\cite{Eto:2004ii} and for domain walls 
  in Ref.~\cite{Eto:2008dm}. For non-Abelian $SO$, $USp$ vortices see
  Ref.~\cite{Eto:2009bg}.}
The moduli space ${\cal M}_k$ contains 
all irreducible $SU(N)$ orbits associated 
with all possible Young tableaux having $k$ boxes, 
as its subspaces which are invariant 
under the action of the spatial rotation. 
In the following, we uniquely determine the metrics for all
irreducible $SU(N)$ orbits. 

The irreducible orbits are all K\"ahler manifolds although generic
$SU(N)$ orbits are not.\footnote{All irreducible
  $SU(N)$ orbits, which are the set of zeros of the holomorphic Killing vector 
  for the spatial rotation, can be obtained as subspaces in ${\cal M}_k$ 
by imposing certain {\it holomorphic} conditions. The latter takes the form (apart from 
    the co-axial condition $\sigma_i =0$)  $B=0$ for baryons which are not in a pure irreducible representation.
       Therefore the K\"ahler
  metrics are induced by these constraints from the K\"ahler metric on
  ${\cal M}_k$.  It is an interesting question if a K\"ahlerian
  coset space in ${\cal M}_k$  always corresponds to an irreducible orbit.}
We shall derive the K\"ahler potentials instead of the metrics directly.

The pair of matrices $(\psi, Z)$ corresponding to generic points on an
orbit is obtained by acting with $SU(N)$ on a specific configuration
$(\psi_0, Z_0)$.
Let us decompose any element $\mathcal{U} \in SU(N)$ as
\beq
\mathcal U = L D U, 
\label{eq:ldu}
\eeq
where $D$ is a diagonal matrix of determinant one and $L$ ($U$) is a
lower (upper) triangular matrix whose diagonal elements are all $1$. 
This is called the LDU decomposition.\footnote{An invertible matrix
  admits an LDU decomposition if and only if all its principal minors
  are non-zero. }
In this case, the matrix $\mathcal{U}$ is a unitary matrix 
$\mathcal{U}\mathcal{U}^\dagger = \mathbf{1}$, 
and hence the matrices $L$, $D$ and $U$ are related by
\beq
U U^\dagger = (LD)^{-1} (LD)^{\dagger-1}.
\label{eq:Cholesky}
\eeq
Therefore, once the matrix $U$ is given, the lower triangular matrix
$LD$ is uniquely determined up to multiplication of diagonal unitary
matrices $u$ as $L D ~\rightarrow~ u L D$. That is, entries of $U$ are
complex coordinates of the flag manifold $SU(N)/U(1)^{N-1}$. 

Let $\psi_0$ and $Z_0$ be matrices of the form given in
Fig.\,\ref{fig:psi_Z} and $m=[m_1,m_2,\cdots,m_{N-1}]$ be the set of
Dynkin labels of the corresponding highest-weight state. 
Since the matrices $\psi_0$ and $Z_0$ satisfy the conditions
\eqref{eq:cond1} and \eqref{eq:cond2}, $LD$ can be always absorbed by 
$g \in GL(k,\mathbb{C})$ and $\tilde{g} \in SL(k,\mathbb{C})$ 
given in Eq.\,\eqref{eq:def_g}
\beq
\psi_0 \, \mathcal U = (\tilde{g} g) \, \psi_0 \, U , \quad 
Z_0 = (\tilde{g} g) Z_0 (\tilde{g} g)^{-1}.
\label{eq:SUNaction}
\eeq
This implies that a pair $(\psi, Z)$ parametrizing the irreducible
$SU(N)$ orbit is given by  
\beq
\psi_{\rm orbit}=
\psi_0 \, U,\quad Z_{\rm orbit} = Z_0, \qquad 
U = 
\begin{pmatrix}
1 & u_{12} & u_{13} & \cdots & u_{1,N}   \\
  & 1      & u_{23} & \cdots & u_{2,N}   \\
  &        & 1      & \ddots & \vdots    \\
  &        &        & \ddots & u_{N-1,N} \\
  &        &        &        & 1
\end{pmatrix}, \quad
u_{ij} \in \C.
\eeq
The vortex state constructed by the latter is obtained as 
\beq
\left|B_{\rm orbit}\right> ~\equiv~ 
\left|B(\psi_{\rm orbit}, Z_{\rm orbit})  \right>
~=~ \hat U \left|B(\psi_0, Z_0) \right> 
~=~ \det g^{-1} \hat{\mathcal U} \left|B(\psi_0, Z_0) \right>,
\eeq
with operators $\hat U$ and $\hat{\mathcal U}$ 
corresponding to $U$ and $\mathcal U$ respectively.

In supersymmetric theories, $\psi$ and $Z$ can be regarded as chiral
superfields. The complex parameters contained in $U$ are also lifted
to chiral superfields and can be regarded as Nambu-Goldstone
zero-modes of $SU(N)/U(1)^{N-1}$.\footnote{The generic K\"ahler
  potential on $SU(N)/U(1)^{N-1}$, which contains $N-1$ free
  parameters (K\"ahler classes), can be obtained from the method of
  supersymmetric non-linear realizations \cite{Bando:1983ab}. 
  When all chiral superfields contain two Nambu-Goldstone scalars as
  in our case, they are called the pure
  realizations. \label{footnote:snlr}}
If $m_i \neq 0$ for all $i=1,\ldots,N-1$, then $SU(N)$ is broken to
the maximal Abelian subgroup (the maximal torus) $U(1)^{N-1}$ and all
the parameters $u_{ij}$ are physical zero modes. 
One can easily check that the dimension of the flag manifold
$SU(N)/U(1)^{N-1}$ counts the degrees of freedom in $U$. 
On the other hand, if $m_i = 0$ for some $i$'s, then the unbroken group
$H$ is enlarged from the maximal torus $U(1)^{N-1}$ to $SU(N)/H$ being
generalized flag manifolds, from which we can further eliminate some of
$u_{ij}$ by using $GL(k,\mathbb C)$.

Since the vortex moduli space ${\cal M}_k$ has an $SU(N)$ isometry, 
the K\"ahler potential for ${\cal M}_k$,
which is a real function of $\sigma_i$ and $B$, 
should be invariant under the $SU(N)$ transformation
\beq
K( \left| B \right> ) = K( \mathcal{\hat U} \left| B \right> ),
\eeq 
where $\left| B \right>$ is the vortex state vector 
satisfying all the constraints \eqref{eq:constraint1}, 
\eqref{eq:constraint2} and \eqref{eq:constraint3}.  
Furthermore, the $\C^\ast$ transformations on the K\"ahler potential 
should be absorbed by the K\"ahler transformations
\beq
K(e^{\lambda} \left| B \right>) ~=~ K(\left| B \right>) + f(\lambda) + \overline{f(\lambda)}, 
\eeq
since the $\C^\ast$ action on $\left| B \right>$ gives 
a physically equivalent state 
$e^{\lambda} \left| B \right> \sim  \left| B \right>$.
Note that this transformation can be absorbed 
only when $\lambda$ is holomorphic in the moduli parameters. 
We can easily show that the function $f(\lambda)$ has 
the following properties
\beq
&f(2 \pi i) + \overline{f(2 \pi i)} ~=~ f(0) + \overline{f(0)},& \\
&f(\lambda_1 + \lambda_2) + \overline{f(\lambda_1 + \lambda_2)} ~=~ f(\lambda_1) + \overline{f(\lambda_1)} + f(\lambda_2) + \overline{f(\lambda_2)}. &
\eeq
From these relations the form of the function $f$ can be determined as
\beq
f(\lambda) + \overline{f(\lambda)} ~=~ r ( \lambda + \bar \lambda ), \hs{10} r \in \R.
\eeq

Now we are ready to derive the K\"ahler potentials for the irreducible
$SU(N)$ orbits.
With the above assumptions, the K\"ahler potential for the $SU(N)$
orbit can be calculated as 
\beq
K(u_{ij}, \bar u_{ij}) &\equiv& K(\left| B_{\rm orbit} \right>) 
~=~ K(\det g^{-1} \, \mathcal{\hat U} \left| B_0 \right>) \notag \\
&=& K(\left| B_0 \right>) - r \log |\det g \,|^2,  
\label{eq:Ku}
\eeq 
where $B_0=B(\psi_0,Z_0)$.
Since the first term of Eq.\,\eqref{eq:Ku} is a constant, it can be
eliminated by a K\"ahler transformation. 
It follows from Eqs.\,\eqref{eq:Cholesky} and \eqref{eq:def_g} that 
\beq
K(u_{ij}, \bar u_{ij}) ~=~ 
-r \log |\det g\,|^2 ~=~ 
r \sum_{l=1}^{N-1} m_l \log \det (U_l U_l^\dagger),
\label{eq:Kahler}
\eeq
where $U_l$ are $l$-by-$N$ minor matrices of $U$ given by
\beq
U_l = 
\begin{pmatrix}
1 & u_{12}  & \cdots & u_{1,l}   & u_{1,l+1} & \cdots & u_{1,N} \\
  & 1      & \ddots & \vdots   & \vdots    &        & \vdots \\
  &        & \ddots & u_{l-1,l}  & \vdots    &        & \vdots \\
  &        &        & 1         & u_{l,l+1} & \cdots & u_{l,N}
\end{pmatrix}.     
\eeq
Note that if $m_l = 0$ for some $l$'s, the dimension of the manifold 
decreases in a way that is consistent with the enhancement of the
symmetry $H$. 

The coefficients $r \, m_l$ of the terms in the K\"ahler potential 
\eqref{eq:Kahler} determine the K\"ahler class of the manifold. 
As noted in the footnote \ref{footnote:snlr} the generic K\"ahler
potential contains $N-1$ free parameters, which is now determined from
the set of Dynkin labels $[m_1,m_2,\cdots, m_{N-1}]$. 
We see that the K\"ahler classes are quantized in integers
multiplied by $r$ which implies that these K\"ahler manifolds are
Hodge. 
This can be expected from the Kodaira theorem stating that Hodge
manifolds are all algebraic varieties, i.e.~they can be embedded into
some projective space ${\mathbb C}P^n$ by holomorphic constraints. 

The overall constant $r$ of the K\"ahler potential cannot be
determined by the above argument based on symmetry. 
It can however be obtained by a concrete computation, for instance,
$k=1$ vortex $(m=[1,0,\cdots,0])$ results  
in Refs.~\cite{Hanany:2004ea,Shifman:2004dr,Eto:2004rz,Gorsky:2004ad} 
\beq
r = \frac{4\pi}{g^2},
\eeq
which matches the result \eqref{eq:kahlerclass} based on the
$D$-brane picture \cite{Hanany:2003hp}. 
It can be also determined from the charge of instantons trapped inside
a vortex \cite{Eto:2004rz}. 

Recently, some of us constructed \cite{Gudnason:2010rm} the
world-sheet action and computed the metrics explicitly from first
principles for the vortices in $SO$,  $USp$ and $SU$ theories,
generalizing the work of
Refs.~\cite{Shifman:2004dr,Gorsky:2004ad}. The systems considered
include the cases of some higher-winding vortices in $U(N)$ and $SO(2N)$
theories: the results found there are in accordance with the general discussion
given here.

\subsection{Examples}

In this subsection we provide two examples with $N=2$ and $N=3$ to 
illustrate the determination of the K\"ahler potentials. 

\subsubsection{$N=2$}
To be concrete, let us take some simple examples for $N=2$.
For simplicity, we first consider the $k=2$ case. 
There are two highest-weight states: the triplet and singlet, for which
$\psi_0$ and $Z_0$ take the form, see Fig.\,\ref{fig:psi_Z}, 
\beq
\def\YGbox{4}
(\psi,Z)_{\Young[0]{2}} 
= \ba{cc|cc} 1 & 0 & 0 & 0 \\ 0 & 0 & 1 & 0 \ea, \quad
(\psi,Z)_{\Young[-0.5]{11}} 
= \ba{cc|cc} 1 & 0 & 0 & 0 \\ 0 & 1 & 0 & 0 \ea.
\eeq
In the former case $SU(2)$ is broken to $U(1)$ and the orbit is
$SU(2)/U(1) \cong \mathbb{C}P^1$.  
Applying Eq.\,\eqref{eq:Kahler}, we obtain the K\"ahler potential for
the Fubini-Study metric on $\C P^1$ 
\beq
K_{N=2} = 2 \, r \log ( 1 + |a|^2 ), \qquad  
U = 
\begin{pmatrix}
1 & a \\ 
0 & 1 
\end{pmatrix}.
\eeq
On the other hand, $SU(2)$ is unbroken in the singlet case. 
Indeed $\psi_0$ is just the unit matrix, so that an arbitrary $SU(2)$ 
transformation can indeed be canceled by $GL(2,\mathbb{C})$.

This can be easily extended to the generic case with $k > 2$. In the
case of $k_1 > k_2$, $SU(2)$ is broken to $U(1)$  
while if $k_1=k_2$, $SU(2)$ is unbroken.
From Eq.\,\eqref{eq:Kahler}, we find the K\"ahler potential 
for the Fubini-Study metric on ${\mathbb C}P^1$ 
for $k_1 > k_2$: 
\beq
K_{N=2} = r \, m_1 \log(1+|a|^2), \quad 
m_1 = k_1 - k_2,
\eeq 
while the orbits are always $\mathbb{C}P^{1}$ for arbitrary $k_1$ and
$k_2$ ($k_1 > k_2$), one can distinguish them by looking at the
K\"ahler class $r m_1 =r (k_1 - k_2)$.  
For instance, one can distinguish two ${\mathbb C}P^1$'s in
Eqs.\,\eqref{eq:k=1} and \eqref{eq:k=2p} for one and two vortices,
respectively.

\subsubsection{$N=3$\label{sec:n3}}
Next, let us study the $N=3$ case.
There are four different types according to the Young tableaux and the 
unbroken groups $H$, see Table \ref{table1}. 
We parametrize the matrix $U$ as
\beq
U = 
\begin{pmatrix}
1 & a & b \\
0 & 1 & c\\
0 & 0 & 1
\end{pmatrix}.
\eeq
The complex parameters $a,b,c$ are (would-be) Nambu-Goldstone
zero-modes associated with $SU(3)\to H$. 
Applying Eq.\,\eqref{eq:Kahler}, we find 
\beq
K_{N=3} = 
r \, m_1 \log\left(1+|a|^2+|b|^2\right) + 
r \, m_2  \log \left(1+|c|^2+|b-ac|^2\right), 
\label{eq:SU(3)/U(1)^3}
\eeq
with $m_1 = k_1 - k_2$ and $m_2 = k_2-k_3$. 
When $m_1 > 0$ and $m_2 > 0$ $(k_1 > k_2 > k_3)$, this represents the
K\"ahler potential for the K\"ahler manifold $SU(3)/U(1)^2$ with a
particular choice of the complex structure \cite{Buchmuller:1985cj}. 
When $m_1>0$ and $m_2 = 0~(k_1 > k_2 = k_3)$, the parameter $c$
disappears from the K\"ahler potential and hence it reduces to 
\beq
K_{m_2=0} = rm_1 \log \left(1+|a|^2+|b|^2\right),
\label{eq:cp2}
\eeq
which is nothing but the K\"ahler potential of 
$\mathbb{C}P^2 \simeq SU(3)/[U(1)\times SU(2)]$. 
When $m_1 = m_2=0~(k_1 = k_2 = k_3)$, $SU(3)$ is unbroken, so that the
orbit is just a point (with a vanishing K\"ahler potential).

\begin{table}
\begin{center}
\begin{tabular}{c|c|c|c|c}
& $k_1 > k_2 > k_3$ & $k_1 > k_2 = k_3$ & $k_1=k_2 > k_3$ & $k_1=k_2=k_3$ \\
\hline
YT & $\Young[-1]{542}$ & $\Young[-1]{522}$ & $\Young[-1]{552}$ & $\Young[-1]{555}$ \\
$H$ & $U(1)^2$ & $U(1)\times SU(2)$ & $U(1)\times SU(2)$ & $SU(3)$
\end{tabular}
\caption{{\small Four different types of $N=3$ coincident vortices.}}
\label{table1}
\end{center}
\end{table}

\subsection{Conjugate orbits \label{sec:conj}}  

Note that in the $SU(3)$ example discussed in the last subsection the replacement
\beq
a\to -c,\quad  b\to ac -b, \quad c\to -a  \label{eq:duality}
\eeq
together with the exchange $m_1 \leftrightarrow m_2$,  leaves invariant the K\"ahler potential
\eqref{eq:SU(3)/U(1)^3}. 
In other words,  irreducible orbits for $m=[m_1,m_2]$ and $m=[m_2,m_1]$ are  identical. 
 In fact,  this is a special case of duality  between two $SU(N)$ conjugate representations, relating the  irreducible orbits for
 $[m_1,m_2,\cdots,m_{N-1}]$ to the one with  $[m_{N-1},m_{N-2},\cdots,m_1]$.   As we are interested here in the motion of the orientational moduli parameters only, 
 it is very reasonable that we find the same K\"ahler metric for a vortex in ${\bf r}$ representation and another in ${\bf r}^{*}$ representation. 
  
Generalization to arbitrary  ($N, k$)   of the mapping (\ref{eq:duality}) leaving the K\"ahler potential invariant is given by  
\beq    [m_{1}, m_{2}, \cdots, m_{N-1}]  & \leftrightarrow  &    [m_{N-1}, m_{N-2}, \cdots, m_{1}] \;, \nonumber \\
     U   & \leftrightarrow  &    E\, (U^{\rm T})^{-1}\, E\;,
\eeq
where $(E)_{i j}=  \delta_{i, N-j+1}$. 

Coming back to the concrete $SU(3)$ examples in Subsection~\ref{sec:n3},   the case with $(k_1,k_2,k_3)=(2,1,0)$ corresponds to 
${\bf 8}$ of $SU(3)$ which of course is self-dual. 
A pair of $(k_1,k_2,k_3)=(3,3,0)$ and $(4,1,1)$ provides a nontrivial 
example of duality between two different irreducible orbits: 
they correspond to  ${\bf 10^{*}}$ and ${\bf 10}$, respectively. 
Finally, the orbits  $(k_1,k_2,k_3)=(5,4,0)$ and $(k_1,k_2,k_3)=(6,2,1)$
belong to the pair of  irreducible representations, 
${\bf 35}^{*}$ and ${\bf 35}$.

 Actually, these examples are 
 special, in the sense that the pairs have the same winding number. This is not necessary.  The equality of the K\"ahler potential (the same effective action)  for a pair of conjugate orbits defined above, holds for pairs of vortices of unequal winding numbers as well, as the above proof does not depend on the winding number, but on the Dynkin labels only. For instance, the $k=1$ vortex in $SU(N)$,    $m =[1,0,\cdots, 0 ]$  (belonging to ${\bf N}$),  has the same K\"ahler potential as the totally antisymmetric vortex of winding number $k=N-1$,
  $m =[0,\cdots, 0,1 ]$. The latter  transforms as ${\bf N}^{*}$.
  
  When the condition $\tfrac{2k}{N} \in  {\mathbb Z}$ is met, it is possible to have pairs of conjugate vortices with the same $k$ (the same tension)  and belonging to conjugate representations, as in the concrete $SU(3)$  examples above.

\section{Summary and outlook}
\label{sec:conclusion}

By using the K\"ahler-quotient construction 
we have investigated the moduli spaces of 
higher-winding BPS non-Abelian vortices in $U(N)$ theory, 
for the purpose of clarifying the transformation properties of 
the points in the moduli under the exact global $SU(N)$ symmetry group. 
In the case of vortices with distinct centers,
the moduli space is basically just the symmetrized direct product 
of those of individual vortices, 
$\left(\mathbb{C}\times \mathbb{C}P^{N-1}\right)^k/\mathfrak{S}_k$.  
It turns out to be a rather nontrivial problem 
to exhibit the group-theoretic properties of the points in the submoduli, 
corresponding  to the vortex solutions with a common center. 
The results found show that they do behave as 
a superposition of various ``vortex states''
corresponding to the irreducible representations, 
appearing in the standard $SU(N)$ decomposition of the products 
of $k$ objects in the fundamental representations (Young tableaux). 

In particular, various ``irreducible $SU(N)$ orbits'' have been identified: 
they correspond to fixed-point sets invariant under the spatial rotation group. These solutions are axially symmetric and they transform according 
to various irreducible representations appearing 
in the decomposition of the direct product.

Although some of our results might be naturally 
expected on general grounds, 
a very suggestive and nontrivial aspect of our findings 
is the fact that the points of the vortex moduli space,
 describing the degenerate set of classical extended field configurations, 
are formally mapped to oscillator ``quantum-state'' vectors, 
endowed with simple $SU(N)$ transformation properties. 
Also, the way the irreducible orbits are embedded 
in the full moduli space appears to be quite nontrivial, and exhibits special features of our vortex systems.
For instance, an irreducible orbit associated with 
a definite type of Young tableau appears only once, 
unlike in the usual decomposition of $k$ distinguishable objects 
in ${\bf N}$. 

We have determined the K\"ahler potential on each of these irreducible orbits.  
Since we have used symmetry only, our K\"ahler 
potential cannot receive any quantum corrections 
except for the overall constant $r$ 
even in non-supersymmetric theories\footnote{ 
The renormalization group flow for $r$  in the case of $k=1$ vortex in 
${\cal N}=2$ $U(N)$ supersymmetric theories was found in 
Refs.~\cite{Hanany:2004ea,Shifman:2004dr}.}. 
The results found agree with some explicit calculations 
made recently by some of us \cite{Gudnason:2010rm}. 

Extension of our considerations to more general situations in $U(N)$ theories 
(question of non-irreducible, general orbits in the vortex moduli space considered here,   or  the metric in the case of semi-local vortices, 
which occur when the number of flavors exceeds the number of colors
\cite{Shifman:2006kd,Eto:2007yv}) remains an open issue. 
A particularly interesting extension would 
however be the study of a more general class of gauge theories, 
such as $SO, USp$ or exceptional groups, 
as the group-theoretic features of our findings 
would manifest themselves better in such wider testing grounds. 
Non-Abelian vortices were constructed in the $G' \times U(1)$
gauge theories with an arbitrary compact Lie group $G'$, 
and the orientational moduli space was found to be $G'/H$ 
with some subgroup $H$ \cite{Eto:2008yi}. 
For instance they are $SO(2N)/U(N)$ and $USp(2N)/U(N)$ 
in the cases of $G'=SO(2N), USp(2N)$. 
The $SO$ and $USp$ non-Abelian vortices and 
their moduli  have been further studied in detail 
in the Refs.~\cite{Ferretti:2007rp,Eto:2008qw,Eto:2009bg,Gudnason:2010jq,Gudnason:2010rm}. 
Especially, $G'$ orbits in the moduli spaces of $SO$ and $USp$
non-Abelian vortices have been studied in Ref.~\cite{Eto:2009bg}.  
Irreducible orbits in these cases may be classified by
(skew-)symmetric Young tableaux. 

Finally, a possible relation to Young tableaux for Yang-Mills
instantons \cite{Nakajima:1999} and its application to the instanton 
counting \cite{Nekrasov:2002qd} may be interesting.
For the instanton counting,
the integration over the instanton moduli space
is reduced to a sum over the Young tableaux,
which correspond to fixed points of the instanton moduli space
under a linear combination of the $SU(N)$ action
and spatial rotations, as in our case of vortices.
Roughly speaking possible vortex counting
should be the half of the instanton counting
since Yang-Mills instantons can stably exist
even in the Higgs phase when they are trapped inside non-Abelian
vortices \cite{Eto:2004rz}.
The partition function of the non-Abelian vortex gas was derived
on a torus and a sphere in Ref.~\cite{Eto:2007aw}
by using a completely different approach of D-brane configurations
and T-duality on it.
A relation with such an approach and the Young tableaux for vortices
developed in this paper appears to be an interesting future
venue to explore.

\section*{Acknowledgments} 

The work of M.E. is supported by Special Postdoctoral Researchers Program at RIKEN. The work of M.N. is supported in part by Grant-in-Aid for Scientific Research
No. 20740141 from the Ministry of Education, Culture, Sports, Science and Technology-Japan.

\begin{appendix}
\section{Constraints on the invariants}\label{appendix:constraints}
In this appendix, we derive the constraints \eqref{eq:constraint1},
\eqref{eq:constraint2} and \eqref{eq:constraint3} from the definition
of the baryons  
\beq
B^{n_1}_{\, r_1}{}^{n_2}_{r_2}{}^{\cdots}_{\cdots}{}^{\, n_k}_{\, r_k}{} 
~\equiv~ 
\epsilon^{i_1 i_2 \cdots i_k} Q^{(n_1)}_{i_1 r_1} Q^{(n_2)}_{i_2 r_2}
\cdots Q^{(n_k)}_{i_k r_k}, \qquad
( Q^{(n)} \equiv Z^n \psi ),
\eeq
and the vortex state vector 
\beq
| B \rangle ~~\equiv
\sum_{{}^{n_1,}_{\hspace{1pt}r_1,}{}^{n_2,}_{\hspace{1pt}r_2,}{}^{\cdots,}_{\cdots,}{}^{n_k}_{\hspace{1pt}r_k}} 
\frac{1}{(n_1! n_2!\cdots n_k!)^{\frac{1}{2}}} \baryon 
\left| n_1,r_1 \right> \otimes 
\left|n_2, r_2 \right> \otimes \cdots \otimes 
\left| n_k, r_k \right>.
\eeq

\begin{enumerate}
\item
Eq.\,\eqref{eq:constraint1} implies that the baryon is anti-symmetric
under the exchange of any pair of indices $(n,r)$.
This can easily be seen from the definition of the baryons
\beq
B^{n_1}_{\, r_1}{}^{\cdots}_{\cdots}{}^{\, n_I}_{\, r_I}{}^{\cdots}_{\cdots}{}^{\, n_J}_{\, r_J}{}^{\cdots}_{\cdots}{}^{\, n_k}_{\, r_k}{}
&=& \phantom{-} \epsilon^{i_1 \cdots i_I \cdots i_J \cdots i_k} Q^{(n_1)}_{i_1 r_1} \cdots Q^{(n_I)}_{i_I r_I} \cdots Q^{(n_J)}_{i_J r_J} \cdots Q^{(n_k)}_{i_k r_k} \notag \\
&=& - \epsilon^{i_1 \cdots i_J \cdots i_I \cdots i_k} Q^{(n_1)}_{i_1 r_1} \cdots Q^{(n_J)}_{i_J r_J} \cdots Q^{(n_I)}_{i_I r_I} \cdots Q^{(n_k)}_{i_k r_k} \notag \\
&=& 
- B^{n_1}_{\, r_1}{}^{\cdots}_{\cdots}{}^{\, n_J}_{\, r_J}{}^{\cdots}_{\cdots}{}^{\, n_I}_{\, r_I}{}^{\cdots}_{\cdots}{}^{\, n_k}_{\, r_k}{}.
\eeq
 
\item 
The annihilation operator $\hat{a}_I$ acts on the state as
\beq
\hat{a}_I | B \rangle &=& \sum \frac{1}{(n_1 ! \cdots (n_I-1) ! \cdots n_k !)^{\frac{1}{2}}} B^{n_1}_{\, r_1}{}^{\cdots}_{\cdots}{}^{\, n_I }_{\, r_I}{}^{\cdots}_{\cdots}{}^{\, n_k}_{\, r_k}{} | n_1, r_1 \rangle \cdots | n_I - 1, r_I \rangle \cdots | n_k, r_k \rangle \notag \\
&=& \sum \frac{1}{(n_1 ! \cdots n_I ! \cdots n_k !)^{\frac{1}{2}}} B^{n_1}_{\, r_1}{}^{\cdots}_{\cdots}{}^{\, n_I+1 }_{\hs{3} r_I \hs{1} }{}^{\cdots}_{\cdots}{}^{\, n_k}_{\, r_k}{} | n_1, r_1 \rangle \cdots | n_I, r_I \rangle \cdots | n_k, r_k \rangle.
\eeq
This means that the baryon is mapped by the operator $\hat{a}_I$ as
\beq
B^{n_1}_{\, r_1}{}^{\cdots}_{\cdots}{}^{\, n_I}_{\, r_I}{}^{\cdots}_{\cdots}{}^{\, n_k}_{\, r_k}{} 
&\mapsto&
B^{n_1}_{\, r_1}{}^{\cdots}_{\cdots}{}^{\, n_I+1 }_{\hs{3} r_I \hs{1} }{}^{\cdots}_{\cdots}{}^{\, n_k}_{\, r_k}{} = \epsilon^{i_1 \cdots j \cdots i_k} Z_{j i_I} Q^{(n_1)}_{i_1 r_1} \cdots Q^{(n_I)}_{i_I r_I} \cdots Q^{(n_k)}_{i_k r_k}.
\eeq
Therefore, we find that the operator 
$\prod_{I=1}^k (\lambda - \hat{a}_I)$ acts on the baryons as 
\beq
B^{n_1}_{\, r_1}{}^{\cdots}_{\cdots}{}^{\, n_k}_{\, r_k}{} 
&\mapsto& \epsilon^{j_1 j_2 \cdots j_k} (\lambda \mathbf 1_k - Z)_{j_1 i_1} \cdots (\lambda \mathbf 1_k - Z)_{j_k i_k} Q^{(n_1)}_{i_1 r_1} \cdots Q^{(n_k)}_{i_k r_k} \notag \\
&=& \det ( \lambda \mathbf 1_k - Z ) B^{n_1}_{\, r_1}{}^{\cdots}_{\cdots}{}^{\, n_k}_{\, r_k}{}.
\eeq
Namely, the vortex state should be an eigenstate of the operator
$\prod_{I=1}^k (\lambda - \hat{a}_I)$  
\beq
\prod_{I=1}^k (\lambda - \hat a_I) | B \rangle = 
\det ( \lambda \mathbf 1_k - Z ) | B \rangle.
\eeq
Comparing the coefficient of $\lambda^i$ on both sides, we obtain the
constraint \eqref{eq:constraint2}. 

\item 
The left hand side of Eq.\,\eqref{eq:constraint3} is
\beq
B^{A_1 \cdots [A_k} B^{B_1 \cdots B_k]} = 
\sum_{i_1, \cdots i_k} \sum_{j_1, \cdots, j_k} 
\epsilon^{i_1 \cdots i_k} \epsilon^{j_1 \cdots j_k}
Q_{i_1}^{A_1} \cdots Q_{i_k}^{[A_k} Q_{j_1}^{B_1} \cdots Q_{j_k}^{B_k]},
\eeq
where $A_i$ and $B_i$ each denote a pair of indices $(n,r)$.
Let us focus on the following part
\beq
\sum_{j_1, \cdots, j_k} \epsilon^{j_1 \cdots j_k} 
Q_{i_k}^{[A_k} Q_{j_1}^{B_1} \cdots Q_{j_k}^{B_k]}. 
\label{eq:A9}
\eeq 
Since the indices $j_1, \cdots , j_k$ are contracted with 
$\epsilon^{j_1 \cdots j_k}$, 
there exist a number $I~(1 \leq I \leq k)$ such that $i_k = j_I$ for
each term in the sum.  
Therefore, all the terms in Eq.\,\eqref{eq:A9} vanish since the
indices $A_k$ and $B_1,\cdots,B_k$ are anti-symmetrized.  
This fact leads to the constraint Eq.\,\eqref{eq:constraint3}. 

\end{enumerate}

\section{A toy metric on the vector space spanned by $\left| B \right>$}
\label{sec:toy}

We have not considered in the main text  the metric for the vector space spanned by 
$\left| B \right>$, introduced in Subsection~\ref{sec:invariants}, for reasons explained at the end of Subsection~\ref{sec:sepa}.   Such a metric would however  induce a natural metric on the vortex moduli space, which is of physical interest.
For instance, one could simply assume the standard inner product 
$\langle B | B \rangle$;    it would induce a metric specified by the
following K\"ahler potential 
\beq
K_{\rm toy} = r \log \left< B | B \right>. \label{eq:toymodel}
\eeq
Note that the equivalence relation \eqref{eq:U1equiv} is realized as
K\"ahler transformations.  
Namely, the moduli space is embedded into the projective space with
suitable constraints \eqref{eq:constraint3}.  
In the case of well-separated vortices $|z_I-z_J|\gg m^{-1}$, we find
that the K\"ahler potential \eqref{eq:toymodel} takes the form 
\beq
 K_{\rm toy}=r\sum_{I=1}^k\left(|z_I|^2+\log |\vec \phi^I|^2\right)
-r \sum_{I,J(\not=I)}
\frac{|\vec \phi^{I\dagger}\cdot 
 \vec \phi^J|^2}{|\vec \phi^I|^2|\vec \phi^J|^2}e^{-|z_I-z_J|^2}
    +\cdots. \label{eq:toy}
\eeq
The first term correctly describes free motion of $k$ vortices while
the second term describes interactions between the vortices.

Unfortunately, the interaction terms do not have the correct form;
terms which behave as $1/|z_I-z_J|^2$ or $K_0(m|z_I-z_I|)$ must be present if
massless or massive modes propagate between vortices, respectively. 
The former is the case of the Hanany-Tong metric \cite{Hanany:2003hp}
(which still does not describe the correct interactions), while the latter
is the case of the correct asymptotic form obtained from the BPS
equations \cite{Fujimori:2010fk}. 

\section{Metrics on $W{\mathbb C}P^2_{(2,1,1)}$ for $k=2$ and
  $N=2$ \label{sec:WCP2}}

In this Appendix we will study some metrics on the intrinsic subspace
$W\mathbb{C}P^2_{(2,1,1)}$ for $k=2$ coincident vortices in the $U(2)$
gauge theory ($N=2$). 
We show that two different metrics on $W\mathbb{C}P^2_{(2,1,1)}$
contain the Fubini-Study metric with the same K\"ahler class on
${\mathbb C}P^1$ at the diagonal edge of Fig.\,\ref{fig:wcp2}. 

For any choice of metric on the moduli space, a subspace specified by
a holomorphic constraint should also be a K\"ahler manifold. 
Its K\"ahler potential must be invariant under the global $SU(2)$ and the
transformation \eqref{eq:phase-trf} as 
\beq
K_{W\mathbb CP^2} ~=~ r f(X) ~\sim~ r \tilde f(X) + {\rm const.} 
\times \log|\phi_i|^2, \hs{10} X ~\equiv~ \frac{|\phi_i|^4}{r |\eta|^2} 
\eeq
with an arbitrary function $f$.
For the Hanany-Tong model, $f(X)$ can be written as 
\cite{Auzzi:2010jt}\footnote{
Here $w$ is identical to that of Eq.\,(32) of the paper presented by
Auzzi-Bolognesi-Shifman \cite{Auzzi:2010jt}. Actually, we can
reproduce the metric Eq.\,(34) in their work from the above potential. 
},
\beq
f(X)=w^2 - \log(1-w^4), \quad 
w^2=\frac{2 X}{1+X+\sqrt{1+6X +X^2}}.
\eeq
For the toy model \eqref{eq:toymodel} in Appendix \ref{sec:toy},  
$f(X)$ can be written as 
\beq
f(X) = r \log\left(1+ r X \right).
\eeq  
These two models have the same behavior
\beq
 f(X)\sim \left\{
\begin{array}{cc}
 \log X+{\rm const}.,& X\gg 1,\\  
{\rm const.}\times X, & X\ll 1.
\end{array}\right. 
\eeq
Since $\log X\simeq 2\log|\phi_i|^2$, they give the usual Fubini-Study
metric on $\mathbb{C}P^1$ with the same K\"ahler class, $2r$, for
$\eta=0$, and they have a conical singularity at $\phi_i=0$. 
These features are not accidental but are guaranteed for any choice of
the moduli space metric, as we show in Section\,\ref{sec:kahler}.

\section{General solution of the linear constraints for $k=3$}\label{appendix:k=3} 

In this section, we consider the general solution of the linear
constraints \eqref{eq:constraint1} and \eqref{eq:constraint2} for
the $k=3$ case. 
We have seen in Section~\ref{sec:sepa} that the solution can be expressed
by the coherent states 
\beq
\left| B \right> ~= \sum_{r_1,r_2,r_3} \tilde B_{r_1 r_2 r_3} \, 
\hat{\mathcal{A}} \Big( \left| z_1, r_1 \right> \otimes 
\left| z_2, r_2 \right> \otimes 
\left| z_3, r_3 \right> \Big).
\label{eq:k=3_coherent}
\eeq
However, this expression is not valid globally on the moduli space  
since the coherent states become linearly dependent when some vortices
coincide $z_i = z_j$.  
In order to derive a globally well-defined expression for the general
solution, let us rewrite the coherent state of
Eq.\,\eqref{eq:k=3_coherent} as 
\beq
\left| B \right> &=& \frac{1}{3!} \sum_{r_1,r_2,r_3} \sum_{\rho \in \mathfrak S_3} {\rm sign}(\rho) \tilde B_{r_1 r_2 r_3} \, \hat \rho \left| z_1 , r_1 \right> \otimes \left| z_2, r_2 \right> \otimes \left| z_3, r_3 \right> \notag \\
&=& \frac{1}{3!} \sum_{r_1,r_2,r_3} \sum_{\rho \in \mathfrak S_3} {\rm sign}(\rho) \tilde B_{r_1 r_2 r_3} \left| z_{\rho(1)}, r_{\rho(1)} \right> \otimes \left| z_{\rho(2)}, r_{\rho(2)} \right> \otimes \left| z_{\rho(3)}, r_{\rho(3)} \right>,
\eeq
where $\hat{\rho}$ is an element of the symmetric group $\mathfrak
S_3$. Defining an operator $\hat{v}$ by
\beq
\hat{v} &\equiv& \exp \left( z_1 \,\hat{a}_1^\dagger + 
z_2 \,\hat{a}_2^\dagger + z_3 \,\hat{a}_3^\dagger \right),
\eeq
and the action of the symmetric group 
\beq
\hat{\rho} \, \hat{v} \, \hat{\rho}^{-1} &\equiv& 
\exp \left( z_1 \hat{a}_{\rho^{-1}(1)}^\dagger + 
z_2 \hat{a}_{\rho^{-1}(2)}^\dagger + z_3 \hat{a}_{\rho^{-1}(3)}^\dagger \right),
\eeq
we can rewrite the state $\left| B \right>$ as
\beq
\left| B \right> = \frac{1}{3!} 
\sum_{r_1,r_2,r_3} \sum_{\rho \in \mathfrak S_3} {\rm sign}(\rho) \,
\tilde{B}_{r_{\rho^{-1}(1)}r_{\rho^{-1}(2)}r_{\rho^{-1}(3)}}\,
\hat{\rho} \, \hat{v} \, \hat{\rho}^{-1} \left| 0, r_1 \right> \otimes  
\left| 0, r_2 \right> \otimes \left| 0, r_3 \right>.
\label{eq:sol_k=3}
\eeq
This means that the solution $\left| B \right>$ is a linear
combination of $3!=6$ states  
$\hat{\rho} \, \hat{v}\,  \hat{\rho}^{-1} \left| 0 ,r_1 \right> \otimes 
\left| 0,r_2 \right> \otimes \left| 0,r_3 \right>$, 
which form a basis of the vector space of states satisfying the
constraint 
\beq
P(\hat{a}_1, \hat{a}_2, \hat{a}_3) 
\left| B \right> = P(z_1, z_2, z_3) \left| B \right>,
\label{eq:constraint_k=3}
\eeq
for all symmetric polynomials $P$. 
However this basis is well-defined only for separated vortices since
the states become degenerate when some vortices coincide. A globally
well-defined basis can however be constructed as follows. 
Let $\left| S ; r_1,r_2,r_3 ; \{z_i\} \right>$ be the state defined by
\beq
\left| S ; r_1,r_2,r_3 ; \{z_i\} \right> \equiv \frac{1}{3!\Delta} 
\sum_{\rho \in \mathfrak S_3} {\rm sign}(\rho) 
\, \hat{\rho} \, \hat{v}\,  \hat{\rho}^{-1} \left| 0,r_1 \right> \otimes 
\left| 0,r_2 \right> \otimes \left| 0,r_3 \right>,
\eeq
where $\Delta$ is the Vandermonde polynomial 
\beq
\Delta(z_1,z_2,z_3) \equiv (z_1-z_2)(z_2-z_3)(z_3-z_1).
\eeq
This state is a solution of the constraint \eqref{eq:constraint_k=3} 
and well-defined even when the vortex centers coincide
\beq
\left| S; r_1,r_2,r_3 ; \{z_i\} \right> ~\rightarrow~ 
\Delta(\hat{a}_1^\dagger,\hat{a}_2^\dagger, \hat{a}_3^\dagger) 
\left| 0,r_1 \right> \otimes 
\left| 0,r_2 \right> \otimes 
\left| 0,r_3 \right>.
\eeq
The other globally well-defined solutions can be constructed by acting
with polynomials of $\hat{a}_i$ on $\left| S; r_1,r_2,r_3 ; \{z_i\} \right>$. 
Note that any polynomial can be decomposed as
\beq
f(\hat{a}_1,\hat{a}_2,\hat{a}_3) =\sum_{i}   g_{i}(\hat{a}_1,\hat{a}_2,\hat{a}_3) \, 
h_{i}(\hat{a}_1,\hat{a}_2,\hat{a}_3), 
\eeq
where $g_{i}$'s  are symmetric polynomials and $h_{i}$ are polynomials satisfying 
\beq
\langle 0 ,r_1| \otimes \langle 0 ,r_2| \otimes \langle 0 ,r_3|\,
h_{i}( \hat{a}_1, \hat{a}_2, \hat{a}_3 ) \,
P ( \hat{a}_1^\dagger, \hat{a}_2^\dagger, \hat{a}_3^\dagger )
= 0.
\label{eq:harmonic}
\eeq
for all symmetric polynomials $P$ (without the constant term). 
Since the state $\left| S; r_1,r_2,r_3 ; \{z_i\}\right>$ satisfies
\beq
g_{i}(\hat{a}_1,\hat{a}_2,\hat{a}_3) \left| S; r_1,r_2,r_3 ; \{z_i\}\right> 
= g_{i}(z_1,z_2,z_3) \left| S; r_1,r_2,r_3 ; \{z_i\} \right>, 
\eeq
a symmetric polynomial $g_{i}(\hat{a}_1,\hat{a}_2,\hat{a}_3)$ does not
create a new state.  
Therefore, it is sufficient to consider the polynomials 
$h_{i}(\hat{a}_1, \hat{a}_2, \hat{a}_3)$ satisfying
Eq.\,\eqref{eq:harmonic}.  
The space of such polynomials $H$ is a $3! = 6$-dimensional vector
space which can be decomposed as 
\beq
H^{(0)} &\ni& S, \\
H^{(1)} &\ni& {\tilde Y}^1 \hat a_1 + {\tilde Y}^2 \hat a_2 + {\tilde Y}^3 \hat a_3, \\
H^{(2)} &\ni&  {\tilde X}^1 (\hat a_2-\hat a_3)^2 + {\tilde X}^2 (\hat a_3-\hat a_1)^2 + {\tilde X}^3 (\hat a_1-\hat a_2)^2, \\
H^{(3)} &\ni& A (\hat a_1 - \hat a_2)(\hat a_2 - \hat a_3)(\hat a_3 - \hat a_1),
\eeq
where $S,{\tilde Y}^i, {\tilde X}^i,A$ are complex numbers satisfying 
\beq
{\tilde Y}^1+{\tilde Y}^2+{\tilde Y}^3=0, \hs{10} {\tilde X}^1+{\tilde X}^2+{\tilde X}^3=0.
\eeq 
The spaces $H^{(i)}$ are closed under the action of the symmetric
group and the decomposition $H = \oplus_i H^{(i)}$ corresponds to the
decomposition of the regular representation of $\mathfrak{S}_3$. 
Acting with the elements of $H^{(i)}$ on $\left| S \right>$, we obtain
the following basis  
\beq
\left| S \right> &\equiv& \sum_{r_1,r_2,r_3} S_{r_1r_2r_3} \left| S;
r_1,r_2,r_3 ; \{z_i\} \right>, \notag \\
\left| Y \right> &\equiv& \sum_{r_1,r_2,r_3} ({\tilde Y}^1_{r_1r_2r_3} \hat a_1
+ {\tilde Y}^2_{r_1r_2r_3} \hat a_2 + {\tilde Y}^3_{r_1r_2r_3} \hat a_3) \left| S;
r_1,r_2,r_3 ; \{z_i\} \right>, \notag \\
\left| X \right> &\equiv& \sum_{r_1,r_2,r_3} \left({\tilde X}^1_{r_1r_2r_3} (\hat
a_2-\hat a_3)^2 + {\tilde X}^2_{r_1r_2r_3} (\hat a_3-\hat a_1)^2 +
{\tilde X}^3_{r_1r_2r_3} (\hat a_1-\hat a_2)^2\right)  \left| S; r_1,r_2,r_3 ; \{z_i\}
\right>, \notag \\
\left| A \right> &\equiv& \sum_{r_1,r_2,r_3} A_{r_1r_2r_3} (\hat a_1 -
\hat a_2)(\hat a_2 - \hat a_3)(\hat a_3 - \hat a_1) \left| S;
r_1,r_2,r_3 ; \{z_i\} \right>,\notag
\eeq
From the anti-symmetry condition 
$\hat{\rho} \left| B \right> = {\rm sign} (\rho) \left| B \right>$, we
find that for all $\rho \in \mathfrak{S}_3$ 
\beq
S_{r_1 r_2 r_3} &=& S_{r_{\rho(1)} r_{\rho(2)} r_{\rho(3)}}, \\ 
{\tilde Y}^i_{r_1 r_2 r_3} &=& {\rm sign} (\rho) {\tilde Y}^{\rho(i)}_{r_{\rho(1)} r_{\rho(2)} r_{\rho(3)}}, \\
{\tilde X}^i_{r_1 r_2 r_3} &=& {\rm sign} (\rho) {\tilde X}^{\rho(i)}_{r_{\rho(1)} r_{\rho(2)} r_{\rho(3)}}, \\
A_{r_1 r_2 r_3} &=& {\rm sign} (\rho) A_{r_{\rho(1)} r_{\rho(2)} r_{\rho(3)}}.
\eeq
These relations imply that the tensors are in the irreducible
representations of $SU(N)$.  
Note that in the coincident limit $z_1=z_2=z_3$, these states reduce
to 
\beq
\left| S \right> &\rightarrow& \sum_{r_1,r_2,r_3} S_{r_1r_2r_3} ( \hat a_1^\dagger - \hat a_2^\dagger ) ( \hat a_2^\dagger - \hat a_3^\dagger ) ( \hat a_3^\dagger - \hat a_1^\dagger ) \left|0,r_1,r_2,r_3 \right>, \notag \\
\left| Y \right> &\rightarrow& \sum_{r_1,r_2,r_3} ({Y}^1_{r_1r_2r_3} (\hat a_2^\dagger-\hat a_3^\dagger)^2 + {Y}^2_{r_1r_2r_3} (\hat a_3^\dagger-\hat a_1^\dagger)^2 + {Y}^3_{r_1r_2r_3} (\hat a_1^\dagger-\hat a_2^\dagger)^2) \left|0,r_1,r_2,r_3 \right>, \notag \\
\left| X \right> &\rightarrow& \sum_{r_1,r_2,r_3} (X^1_{r_1r_2r_3} \hat a_1^\dagger + X^2_{r_1r_2r_3} \hat a_2^\dagger + X^3_{r_1r_2r_3} \hat a_3^\dagger) \left|0,r_1,r_2,r_3 \right>, \notag \\
\left| A \right> &\rightarrow& \sum_{r_1,r_2,r_3} A_{r_1r_2r_3} \left|0,r_1,r_2,r_3 \right>, \notag 
\eeq
where 
$\left|0,r_1,r_2,r_3 \right> = \left|0,r_1 \right> \otimes 
\left|0,r_2 \right> \otimes \left|0,r_3 \right>$, and 
\begin{align}  Y^1 &\equiv  {\tilde Y}^{2}- {\tilde Y}^{3}, \qquad  & Y^2 & \equiv   {\tilde Y}^{3}- {\tilde Y}^{1}, \qquad  & Y^3 & \equiv  {\tilde Y}^{1}- {\tilde Y}^{2}\;; \nonumber \\
 X^1&  \equiv   -6 \, ( {\tilde X}^{2}- {\tilde X}^{3}), \qquad  & X^2 &\equiv    -6\, ({\tilde X}^{3}- {\tilde X}^{1}), \qquad  & X^3&\equiv   -6 \,( {\tilde X}^{1}- {\tilde X}^{2})\;. 
 \end{align}
By rewriting the solution \eqref{eq:sol_k=3} as a linear combination
of these states, we obtain the globally well-defined general solution
to the linear constraints. 

\end{appendix}


\end{document}